\documentclass[%
 reprint,
 amsmath,amssymb,
 aps,
prb,
]{revtex4-2}

\usepackage{graphicx}
\usepackage{color}
\usepackage{dcolumn}
\usepackage{bm}
\usepackage{hyperref}
\hypersetup{%
 hidelinks,%
 setpagesize=false,%
 bookmarksnumbered=true,%
 colorlinks=false,%
}

\begin{document}

\preprint{JPSJ, Manuscript No. 70967}

\title{ Thermoelectric Effect in Kagome Lattice Enhanced at Van Hove Singularities }

\author{Kaiki Shibata}
 \email{kaikishibata@cphys.s.kanazawa-u.ac.jp}
\affiliation{%
 Graduate School of Natural Science and Technology, Kanazawa University, Kanazawa 920-1192, Japan
}%


\author{Naoya Yamaguchi}
\author{Hikaru Sawahata}
\author{Fumiyuki Ishii}
\email{ishii@cphys.s.kanazawa-u.ac.jp}
\affiliation{
 Nanomaterials Research Institute (NanoMaRi), Kanazawa University, Kanazawa 920-1192, Japan
}%

\date{\today}

\begin{abstract}
We performed first-principles calculations using density functional theory on a kagome lattice model with a chiral spin state as a representative example demonstrating significant longitudinal and transverse thermoelectric properties.
The results revealed that the saddle-point-type van Hove singularity (VHS) enhances thermoelectric effects. 
The longitudinal thermoelectric conductivity $\alpha_{xx}$ was large at the chemical potentials tuned close to the band at the symmetry points K (lower band edge), $\Gamma$ (upper band edge), and M (saddle point), where the VHSs of the density of states were at the corresponding band energies.
The transverse thermoelectric conductivity $\alpha_{xy}$ was large at the chemical potential of the saddle-point-type VHS. 
A large anomalous Nernst coefficient of approximately 10 $\mu$V/K at 50 K was expected.
\end{abstract}

\maketitle


\section{Introduction}
\label{section_introduction}

The thermoelectric effect is a phenomenon in which a temperature gradient generates an electric field. 
Seebeck and Nernst effects are two types of thermoelectric effects 
in which the electric field is parallel and perpendicular to the temperature gradient, respectively. 
The proportionality coefficients between the temperature gradient $\nabla T$ and the electric field ($E$) 
are called the Seebeck coefficient ($S$)
and the Nernst coefficient ($N$) , respectively, and 
are defined as follows: 
\begin{align}
S &\equiv \frac{E_x}{(\nabla T)_x} = \frac{S_0 + \theta_H N_0}{1 + \theta_H^2} , 
\label{eq_def_seebeck} \\
N &\equiv \frac{E_x}{(\nabla T)_y} = \frac{N_0 - \theta_H S_0}{1 + \theta_H^2} ,  
\label{eq_def_nernst}
\end{align}
where $S_0$ = $\alpha_{xx} / \sigma_{xx}$, $N_0$ = $\alpha_{xy} / \sigma_{xx}$, $\theta_H$ = $\sigma_{xy} / \sigma_{xx}$, 
$\sigma_{xx}$, $\sigma_{xy}$, $\alpha_{xx}$, and $\alpha_{xy}$ are 
the pure Seebeck coefficient, pure Nernst coefficient, Hall angle, 
electrical conductivity, Hall conductivity, longitudinal thermoelectric conductivity, 
and transverse thermoelectric conductivity, respectively 
\cite{Mizuta_JPS_Conf_Proc_5_011023_2015, Mizuta_Sci_Rep_6_28076_2016}.

As the thermoelectric effect can potentially be applied for effectively utilizing waste heat, 
elucidating the origins of large $S$ and $N$ values can aid the development of high-performance thermoelectric devices. 
Equations \eqref{eq_def_seebeck} and \eqref{eq_def_nernst} show that 
for a constant temperature gradient $\nabla T$, 
the larger the magnitudes of $S$ and $N$, 
the larger the magnitude of the electric field $E$. 
The numerators of Eqs. \eqref{eq_def_seebeck} and \eqref{eq_def_nernst} indicate that 
large magnitudes of $\alpha_{xx}$ and $\alpha_{xy}$ result in large values of $|S|$ and $|N|$. 
Furthermore, 
$\alpha_{ij}$ is calculated from $\sigma_{ij}$ 
as follows: $\alpha_{ij} = - 1/|e| \int d\epsilon ~ \sigma_{ij} ~ (\epsilon - \mu)/T ~ 
( - \partial f / \partial \epsilon ) $. 
Based on the semiclassical Boltzmann transport theory, 
$\sigma_{xx}$ is given by 
$\sigma_{xx}$ = $e^2 \int d\epsilon \: \Sigma_{xx} (\epsilon) (-\partial f / \partial \epsilon)$, 
where $e$, $\epsilon$, $\mu$, $T$, $f$, and $\Sigma_{xx}$ are the elementary charge, energy, 
chemical potential, absolute temperature, Fermi-Dirac distribution function, and 
transport distribution function, respectively 
\cite{Mahan_Proc_Natl_Acad_Sci_USA_93_7439_1996}. 
Moreover, 
$\Sigma_{xx} (\epsilon)$ is  defined as  
$\Sigma_{xx} (\epsilon)$ = $\sum_{\bm{k}} v_x^2 (\bm{k})  \tau (\bm{ k}) \delta (\epsilon - \epsilon (\bm{k}))$, 
where $\bm{k}$, $v_x$, $\tau$, and $\delta$ are 
the wavenumber vector, group velocity in the $x$ direction, relaxation time, and Dirac delta function, respectively 
\cite{Mahan_Proc_Natl_Acad_Sci_USA_93_7439_1996}. 
If the relaxation time $\tau$ is treated as a constant, $\Sigma_{xx}$ is written as  
$\Sigma_{xx} (\epsilon)$ = $\tau v_x^2 (\epsilon) D(\epsilon)$, 
where 
$v_x^2 (\epsilon)$ = $\sum_{\bm{k}} v_x^2 (\bm{k}) \delta ( \epsilon - \epsilon (\bm{k}) )$ $/$ 
$\sum_{\bm{k}} \delta ( \epsilon - \epsilon (\bm{k}) )$, and the density of states (DOS) 
$D(\epsilon) = \sum_{\bm{k}} \delta (\epsilon - \epsilon (\bm{k}))$. 
In such a case, if $D(\epsilon)$ is large, we can assume that 
$|\alpha_{xx}|$ is large. 
In contrast to $\sigma_{xx}$, the anomalous Hall conductivity $\sigma_{xy}$ is expressed  
using the wavenumber $\bm{k}$-dependent Berry curvature $\Omega_{n, z} (\bm{k})$ 
as follows: 
$\sigma_{xy}$ = $e^2/\hbar \sum_n \int d^d k / (2\pi)^d ~ \Omega_{n,z} (\bm{k}) f(\epsilon_n (\bm{k}), \mu, T)$, 
where $n$ and $d$ are the band index and the dimension of the system, 
respectively. 
Subsequently, we define the energy-dependent Berry curvature $\Omega_z (\epsilon)$ as 
$\Omega_{z} (\epsilon)$ = 
$\sum_{n, \bm{k}} \Omega_{n,z} (\bm{k}) \delta (\epsilon - \epsilon_n (\bm{k}))$  $/$  
$\sum_{n, \bm{k}} \delta (\epsilon - \epsilon_n (\bm{k}))$. 
In this case, $\sigma_{xy}$ is written as 
$\sigma_{xy}$ = $(e^2 / \hbar)$ $(1/V)$ $\int d\epsilon ~ \Omega_z (\epsilon) D(\epsilon) f(\epsilon, \mu, T)$. 
This shows that 
a large $D (\epsilon)$ is expected to yield a large $|\alpha_{xy}|$.
Studies have reported that 
in the three-dimensional (3D) systems with nodal lines, Co$_3$Sn$_2$S$_2$, Co$_2$MnGa, and Fe$_3$Al, 
large $|\alpha_{xy}|$ were obtained at the energies of large $D(\epsilon)$ on the nodal lines 
\cite{Minami_Phys_Rev_B_102_205128_2020}.

We focused on the van Hove singularities (VHSs), 
which are the singularities of the DOS 
\cite{Van_Hove_Phys_Rev_89_1189_1953, book_Grosso_2014}, 
as the origin of large values of $S$ and $N$.  
The DOS is expressed as follows: 
$D(\epsilon)$ = $\int \frac{V}{(2\pi)^3} \frac{dA}{|\nabla_{\bm{k}} \epsilon (\bm{k})|}$ 
as the integral over the isoenergetic surface $A$ with energy $\epsilon$ in $k$-space. 
This equation shows that the DOS has singularities at the critical points 
satisfying $\nabla_{\bm{k}} \epsilon (\bm{k})$ = 0. 
These singularities are the VHSs. Near the VHSs, 
the energy $\epsilon$ can be expanded into a quadratic form 
as a function of the wavenumber vector $\bm{k}$. 
We assumed that $\epsilon (\boldsymbol{k})$ is given by 
$\epsilon (\bm{k})$ = $\epsilon_c 
\pm \frac{\hbar^2}{2m_x} k_x^2 \pm \frac{\hbar^2}{2m_y} k_y^2 \pm \frac{\hbar^2}{2m_z} k_z^2$, 
where $\epsilon_c$ is a constant and $m_x$, $m_y$, and $m_z$ are effective masses. 
In this case, the VHSs exist at $\boldsymbol{k}$ = 0. 
When the signs of this equation are all negative such as 
$\epsilon(\bm{k})$ = $\epsilon_c - \frac{\hbar^2}{2m_x} k_x^2 - \frac{\hbar^2}{2m_y} k_y^2 
- \frac{\hbar^2}{2m_z} k_z^2$, 
$\epsilon(\bm{k})$ is a maximum at $\bm{k}$ = 0. 
Meanwhile, when the signs are all positive such as  
$\epsilon(\bm{k})$ = $\epsilon_c + \frac{\hbar^2}{2m_x} k_x^2 + \frac{\hbar^2}{2m_y} k_y^2 
+ \frac{\hbar^2}{2m_z} k_z^2$, 
$\epsilon (\bm{k}$) is a minimum at $\bm{k}$ = 0. 
Otherwise, $\epsilon (\bm{k})$ is a saddle point at $\bm{k}$ = 0. 
Thus, VHSs are of three types: maximum, minimum, and saddle-point types. 
In two-dimensional (2D) systems, 
near the saddle-point-type VHS $\epsilon_s$, the DOS is expressed as follows: 
$D(\epsilon)$ $\simeq$ $- A \frac{\sqrt{m_x m_y}}{\pi^2 \hbar^2} \ln{|\epsilon - \epsilon_s|}$ 
\cite{Van_Hove_Phys_Rev_89_1189_1953, book_Grosso_2014},  
which shows that $D(\epsilon)$ diverges at $\epsilon$ = $\epsilon_s$. 
Thus, large $|\alpha_{xx}|$ and $|\alpha_{xy}|$ are expected at the saddle-point-type VHS. 
Previous studies on 2D systems such as 
Re$X_2$ ($X$ = S, Se, Te) \cite{Verzola_Mater_Today_Commun_33_104468_2022}, 
FeCl$_2$ \cite{Syariati_APL_Materials_8_041105_2020}, 
and Fe$_3$GeTe$_2$ \cite{Xu_Nano_Lett_19_8250_2019}  
have reported large Seebeck coefficients.
Similarly, previous studies on 2D systems such as 
FeCl$_2$ \cite{Syariati_APL_Materials_8_041105_2020}, 
Fe$_3$GeTe$_2$ \cite{Xu_Nano_Lett_19_8250_2019}, 
and CrTe$_2$ \cite{Yang_Phys_Rev_B_103_024436_2021} 
have reported large anomalous Nernst coefficients. 
However, there have been no discussions regarding VHSs. 
Therefore, this study investigated VHSs as the origin of the enhancement in $S$ and $N$.

This study showed that the saddle-point type VHS yields large thermoelectric coefficients.  
To investigate this, we performed first-principles calculations based on the density functional theory (DFT) 
on the kagome lattice model with a chiral spin state. 
The effects of the DOS and Berry curvatures 
on $\alpha_{xx}$ and $\alpha_{xy}$ were investigated. 
Assuming a constant relaxation time, 
$|\alpha_{xx}|$ was found to be the maximum at the saddle-point-type VHS.  
Moreover, upon assuming 
that the intrinsic contribution was dominant 
in the anomalous Hall conductivity $\sigma_{xy}$, 
$|\alpha_{xy}|$ was the largest at the saddle-point-type VHS.  
Therefore, 
the saddle-point-type VHS was found to yield large thermoelectric coefficients in 2D magnetic materials.

\section{Theory}


%
Using the constant relaxation time approximation, 
the 2D electrical conductivity $\sigma_{xx}$ is given by  
\cite{
Minami_Appl_Phys_Lett_113_032403_2018, 
Syariati_APL_Materials_8_041105_2020}: 
\begin{align}
\sigma_{xx} (T, \mu) = e^2 \tau \sum_n \!\! \int \!\! \frac{d^2 k}{(2 \pi)^2}  v_{n,x}^2 (\bm{k}) \!\!\!
\left. \left(- \frac{\partial f (\epsilon, \mu, T)}{\partial \epsilon} \right) \right|_{\epsilon = \epsilon_{n} (\bm{k})}.  
\label{eq_sigmaxx}
\end{align} 
The 2D anomalous Hall conductivity $\sigma_{xy}$ induced by the Berry curvature 
$\bm{\Omega}_n (\bm{k})$ = $-i \langle \nabla_{\boldsymbol{k}} u_n (\bm{k}) | \times | \nabla_{\boldsymbol{k}} u_n (\bm{k}) \rangle$ 
is written as   
\begin{align}
\sigma_{xy} (T, \mu) =  \frac{e^2}{h} \sum_n 
\int \frac{d^2k}{2 \pi}~ \Omega_{n, z} (\bm{k}) f(\epsilon_{n} (\bm{k}), \mu, T) ,  
\label{eq_sigmaxy_berry}
\end{align} 
where 
$u_n (\bm{k})$ is the periodic part of the Bloch function  
\cite{
Xiao_Rev_Mod_Phys_82_1959_2010, 
Nagaosa_Rev_Mod_Phys_82_1539_2010}.


The thermoelectric conductivity $\alpha_{ij}$ is calculated from $\sigma_{ij}$ as follows ($i$, $j$ = $x$ or $y$) 
\cite{Mizuta_JPS_Conf_Proc_5_011023_2015, Mizuta_Sci_Rep_6_28076_2016}:
\begin{align}
\alpha_{ij} (T, \mu) = - \frac{1}{|e|} \int d\epsilon ~ \sigma_{ij} (T=0, \epsilon) \frac{\epsilon - \mu}{T} 
\left( - \frac{\partial f (\epsilon,\mu,T)}{\partial \epsilon} \right). 
\label{eq_alpha_integ_sigma}
\end{align}
To understand the relationship between the DOS $D(\mu)$ and $\alpha_{ij}$, 
the Mott relation was used to express $\alpha_{ij}$ in terms of $D(\mu)$ 
as follows 
(refer to Appendix \ref{section_Berry_energy} for the derivation): 
\begin{align}
\alpha_{xx}(T, \mu) &= - \frac{\pi^2 |e| k_B^2 T \tau}{3 V} \frac{d}{d\mu} \Bigl( D(\mu) v_x^2 (\mu) \Bigr)  ,   
\label{eq_alpha_xx} \\
\alpha_{xy} (T, \mu) &= - \frac{\pi^2  |e| k_B^2 T}{3 \hbar V}  D(\mu) \Omega_z (\mu) ,  
\label{eq_alpha_dos_berry}
\end{align}
where $k_B$ is the Boltzmann constant.  
For 2D systems, since the DOS diverges at the saddle-point-type VHS 
\cite{Van_Hove_Phys_Rev_89_1189_1953, book_Grosso_2014}, 
$|D(\mu)|$ and $|dD(\mu)/d\mu|$ are large. 
Therefore, large $|\alpha_{xx}|$ and $|\alpha_{xy}|$ are expected. 
At the VHSs, the Mott relation is violated for $|\alpha_{xx}|$ and $|\alpha_{xy}|$ (refer to Appendix \ref{section_Mott_relation_VHS} for more details). However, the tendency in the chemical-potential dependence of the $|\alpha_{xx}|$ and $|\alpha_{xy}|$ at the VHSs can be explained by Eqs. \eqref{eq_alpha_xx} and \eqref{eq_alpha_dos_berry}. 


\section{Computational Details}

First, we performed first-principles calculations based on the density functional theory (DFT) 
using OpenMX (version 3.9) 
\cite{
Ozaki_Phys_Rev_B_67_155108_2003, 
Ozaki_Phys_Rev_B_69_195113_2004, 
Ozaki_Phys_Rev_B_72_045121_2005, 
Lejaeghere_Science_351_aad3000_2016}
to obtain eigen-energy $\epsilon (\boldsymbol{k})$ and Bloch wave function $\psi_{\boldsymbol{k}}$. 
To calculate a chiral spin state, we used non-collinear DFT 
with two-component spinor wave functions for Kohn-Sham-Bloch orbitals 
\cite{vonBarth_J_Phys_C_5_1629_1972, Kubler_J_Phys_F_18_469_1988}. 
The Kohn-Sham equation for the electrons was solved by the self-consistent field (SCF) method. 
The exchange-correlation potential was approximated by the generalized gradient approximation 
(GGA) method \cite{Perdew_Phys_Rev_Lett_77_3865_1996}. 
In OpenMX, core electrons were replaced by pseudopotentials. 
We used pseudo atomic orbitals as the basis function of the wave function. 
In this study, we constructed a kagome lattice model with a chiral spin state using hydrogen atoms. 
Two s-orbitals and one p-orbital were prepared as s2p1. 
The cutoff radius was set to 6.0 Bohr. 
The SCF calculations were performed by discretizing the first Brillouin zone into a 30 $\times$ 30 $\times$ 1 mesh. 
Furthermore, the energy cutoff for the real-space numerical integrations and the solution of Poisson's equation was set to 210 Ry.

Next, the electrical conductivity $\sigma_{xx}$ and anomalous Hall conductivity $\sigma_{xy}$ were calculated 
using $\epsilon (\boldsymbol{k})$ and $\psi_{\boldsymbol{k}}$. 
In this calculation, we used the rigid band approximation, 
which assumes that the bands are invariant with respect to changes in chemical potential $\mu$. 
For $\sigma_{xx}$, we obtained the values by using the semiclassical Boltzmann equation based on the semiclassical theory 
using Wannier90 
\cite{
Arash_Comput_Phys_Commun_178_685_2008, 
Arash_Comput_Phys_Commun_185_2309_2014, 
Weng_Phys_Rev_B_79_235118_2009, 
Pizzi_Comput_Phys_Commun_185_422_2014}. 
In this case, $\sigma_{xx}$ = $e^2 \int d\epsilon \: \Sigma_{xx} (\epsilon) (-\partial f / \partial \epsilon)$,  
where $\Sigma_{xx} (\epsilon)$ is the transport distribution function: 
$\Sigma_{xx} (\epsilon)$ = $\sum_{\bm{k}} v_x^2 (\bm{k})  \tau (\bm{ k}) \delta (\epsilon - \epsilon (\bm{k}))$. 
For the relaxation time $\tau(\boldsymbol{k})$, we considered the constant relaxation time approximation, 
i.e., the energy-independent relaxation time $\tau$. 
In this case, we set $\tau$ = 10 and 100 fs because of the relationship with $\sigma_{xy}$ 
(refer to Appendix \ref{section_relaxation_time} for details). 
The velocity $v_x(\boldsymbol{k})$ was obtained from the Wannier function, 
which is the complete set of orthogonal functions, and the Fourier transform of the Bloch function. 
We used Wannier90 
\cite{
Arash_Comput_Phys_Commun_178_685_2008, 
Arash_Comput_Phys_Commun_185_2309_2014, 
Weng_Phys_Rev_B_79_235118_2009, 
Pizzi_Comput_Phys_Commun_185_422_2014} 
to construct maximally-localized Wannier functions (MLWFs) for the valence bands. 
Velocity $v_x$ was obtained from the Hamiltonian based on the MLWFs. 
Moreover, $\sigma_{xy}$ was obtained from the Berry curvature, 
considering the case where the intrinsic contribution was dominant 
\cite{Onoda_Phys_Rev_B_77_165103_2008} 
(refer to Appendix \ref{section_relaxation_time} for details). 
The Berry curvature was calculated using the method proposed by Fukui, Hatsugai, and Suzuki 
\cite{
Fukui_J_Phys_Soc_Jpn_74_1674_2005}. 
The method was implemented in OpenMX 
\cite{
Sawahata_Jpn_J_Appl_Phys_57_030309_2018, 
Sawahata_Phys_Rev_B_107_024404_2023}.

Figure \ref{fig_kagome_lattice_structure}(a) in Appendix \ref{sec_computational_model} 
shows the kagome lattice composed of the hydrogen atoms 
used in this calculation. 
The positions of the atoms were fixed. 
The unit cell vectors were $a_1$ = $a_2$ = 6.60 \AA, and the angle between $\bm{a}_1$ and $\bm{a}_2$ was  = 120$^\circ$. 
A slab model with a sufficient vacuum layer ($a_3$ = 100 \AA) was used to eliminate the interactions between periodic images.  
To calculate the case of a chiral spin state with an azimuthal angle of 70$^\circ$, 
we used the penalty function \cite{Kurz_Phys_Rev_B_69_024415_2004} to fix the spin orientations. 
Figure \ref{fig_kagome_lattice_structure}(b) in Appendix \ref{sec_computational_model} 
represents the $k$-space. 
The reciprocal lattice vectors were $\bm{b}_1$, $\bm{b}_2$, and $\bm{b}_3$. 
The fractional coordinates of the reciprocal lattice space corresponding to the $\Gamma$, M, and K points were 
(0, 0, 0), (1/2, 0, 0), and (1/3, 1/3, 0), respectively. 
In these coordinates, $2\pi/a$ is set to 1, where $a$ is the length of the unit cell vector.

\section{Results and Discussion}

\begin{figure*}[htbp]
 \begin{center}
  \includegraphics[width=0.9\linewidth]{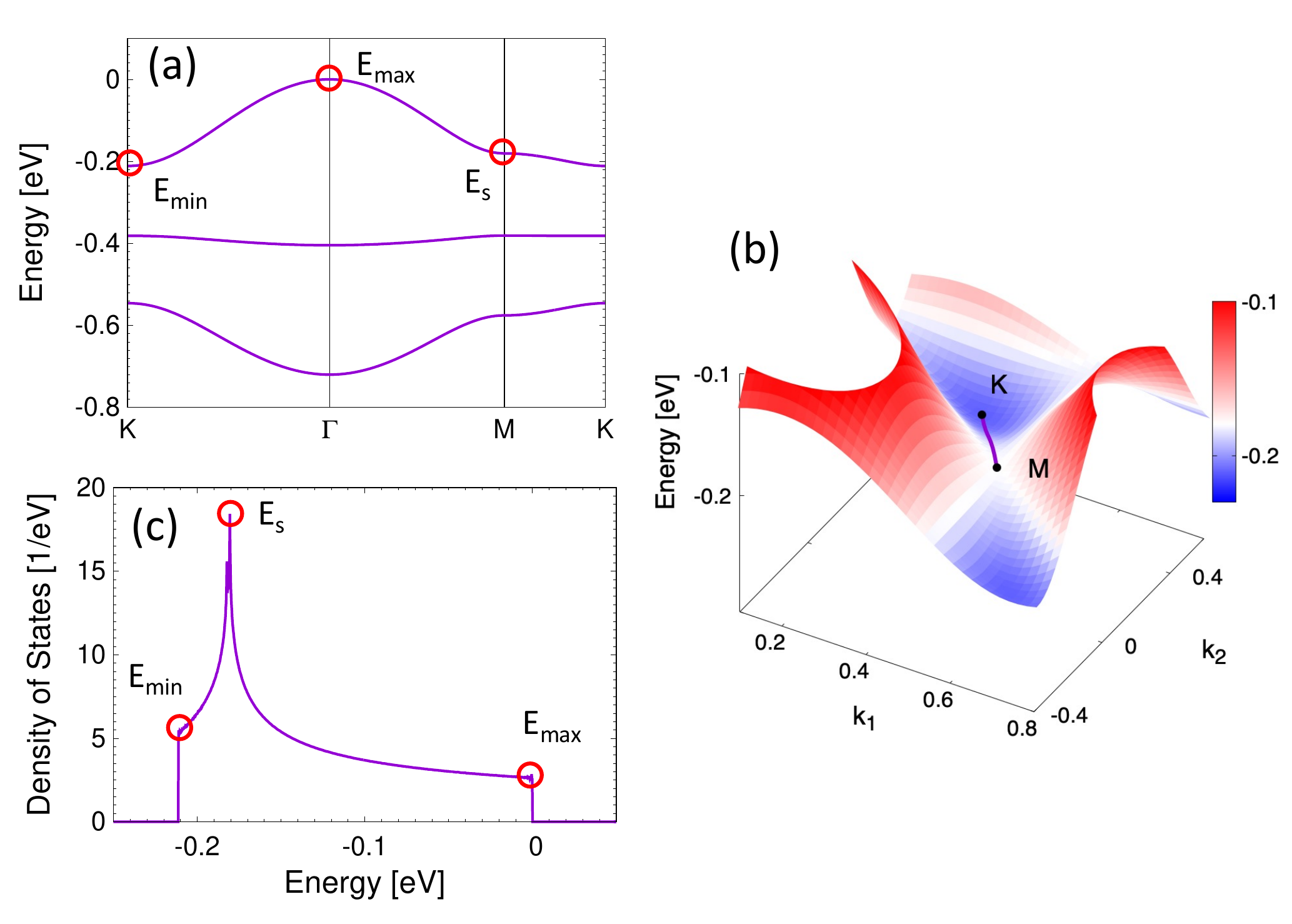}
 \end{center}
 \caption{(Color)
 (a) 
 Band structure of the valence bands of the kagome lattice with a chiral spin state. 
 The red circles represent the energies $E_{\mathrm{max}}$, $E_s$, and $E_{\mathrm{min}}$ 
 at the $\Gamma$(0, 0, 0), M(1/2, 0, 0), and K(1/3, 1/3, 0) points, respectively. 
 These points are expressed as fractional coordinates of the reciprocal lattice space with $2\pi/a$ as 1, 
 where $a$ is the length of the unit cell vector.
 We set $E_{\mathrm{max}}$ = 0 eV. 
 In this case, $E_s$ = $-0.181$ eV and $E_{\mathrm{min}}$ = $-0.211$ eV. 
 (b) 
 Distribution of energy eigenvalues in $k$-space for the top band of the valence bands. 
 The $z$-axis represents energy eigenvalues. 
 The contours are drawn on a linear scale from $E_{\mathrm{min}}$ to $-0.1$ eV.  
 The plotted energies increase from blue to red. 
 Additionally, $k_1$ and $k_2$ are fractional coordinates of the reciprocal lattice space with $2\pi/a$ as 1. 
 The line from the M point to the K point is the $k$-path of M-K in the band structure in (a).
 (c) Density of states of the top band of the valence bands. 
  The maximum, saddle, and minimum-point-type VHSs are obtained at 
 $E_{\mathrm{max}}$, $E_s$, and $E_{\mathrm{min}}$, respectively.
 }
 \label{fig_kagome_lattice_band_dos}
\end{figure*}
\begin{figure*}[htbp]
 \begin{center}
  \includegraphics[width=\linewidth]{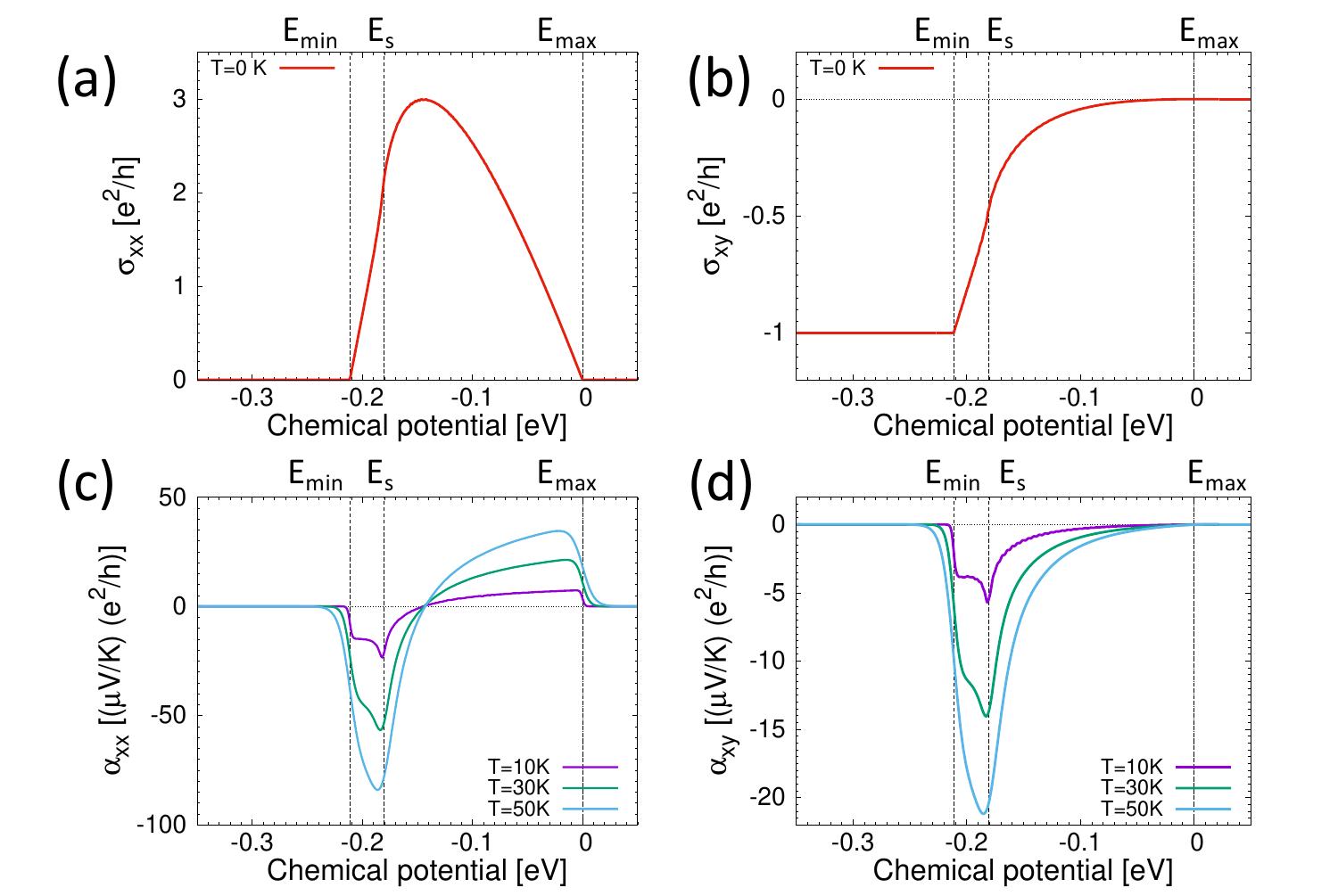}
 \end{center}
 \caption{(Color)
 Chemical potential dependence of 
 (a) electrical conductivity $\sigma_{xx}$, 
 (b) anomalous Hall conductivity $\sigma_{xy}$,  
 (c) longitudinal thermoelectric conductivity $\alpha_{xx}$  
 and (d) transverse thermoelectric conductivity $\alpha_{xy}$. 
 Here, $\sigma_{xx}$ and $\sigma_{xy}$ correspond to $T$ = 0 K, 
 and $\alpha_{xx}$ and $\alpha_{xy}$ correspond to $T$ = 10, 30, and 50 K. 
 The relaxation time $\tau$ used to calculate $\sigma_{xx}$ was approximated as a constant of 10 fs. 
 Furthermore, $E_{\mathrm{min}}$, $E_s$, and $E_{\mathrm{max}}$ correspond to the energies in Fig. \ref{fig_kagome_lattice_band_dos}. 
 }
 \label{fig_kagome_lattice_sigma_alpha}
\end{figure*}
\begin{figure*}[htbp]
 \begin{center}
  \includegraphics[width=\linewidth]{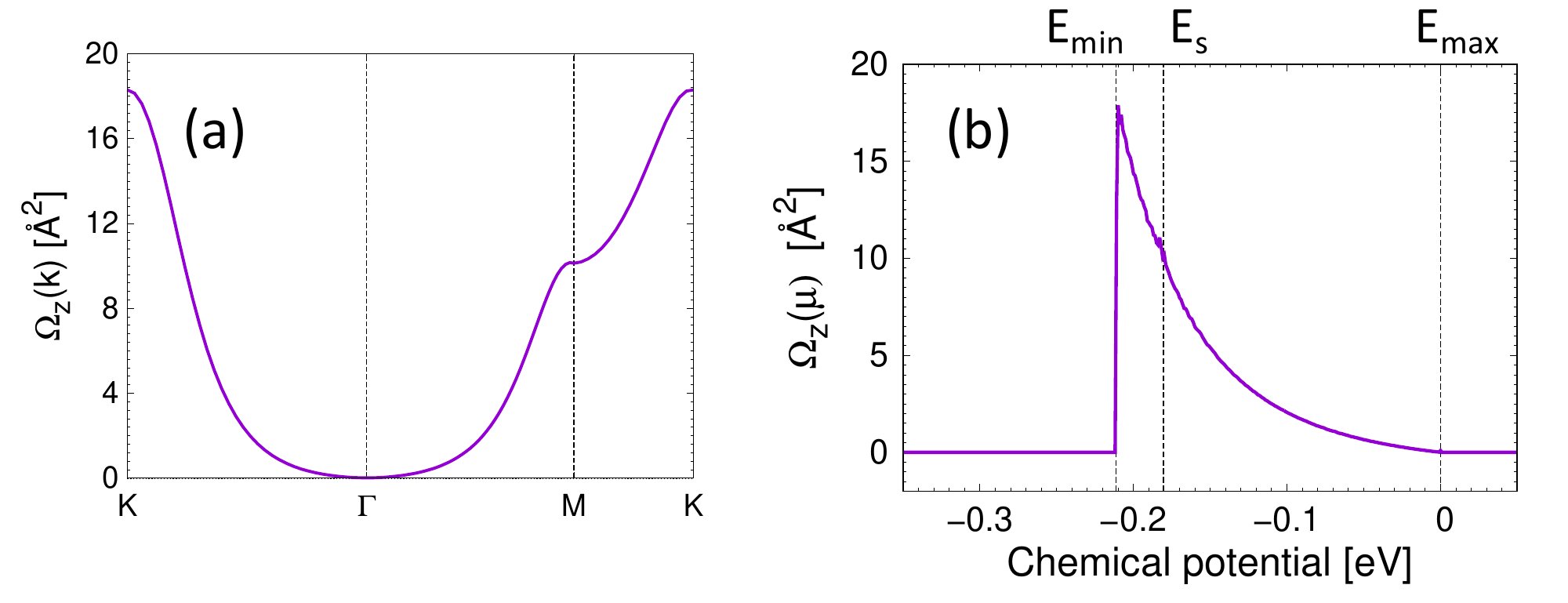}
 \end{center}
 \caption{(Color online)
 (a) Berry curvature $\Omega_z (\boldsymbol{k})$ in $k$-space on the $k$-path of the band structure. 
 (b) Berry curvature $\Omega_z (\mu)$ depending on the chemical potential $\mu$ calculated using Eq. \eqref{eq_Berry_curvature_energy} 
 in Appendix \ref{section_Berry_energy}. 
 Both are the Berry curvatures of the top band of the valence bands. 
 Additionally, $E_{\mathrm{min}}$, $E_s$, and $E_{\mathrm{max}}$ correspond to the energies in Fig. \ref{fig_kagome_lattice_band_dos}. 
 }
 \label{fig_kagome_lattice_berry}
\end{figure*}

\subsection{Electronic Structure}

Figure \ref{fig_kagome_lattice_band_dos}(a) shows the calculated valence band structure. 
Three valence bands were obtained as the number of electrons per unit cell was three. 
We considered hole-doped cases with less than one hole. 
The Fermi level was within the bandwidth of the top band of the valence bands. 
The energy eigenvalues $\epsilon (\boldsymbol{k})$ are maximum, saddle point, and minimum 
at the $\Gamma$, M, and K points, respectively, 
as shown in Fig. \ref{fig_kagome_lattice_band_dos}(b). 
Figure \ref{fig_kagome_lattice_band_dos}(b) shows the distribution of energy eigenvalues in $k$-space. 
In the following text, 
the energies crossing the band at the $\Gamma$, M, and K points are denoted as 
$E_{\mathrm{max}}$, $E_s$, and $E_{\mathrm{min}}$, respectively. 
Figure \ref{fig_kagome_lattice_band_dos}(c) shows that in the energy range 
$E_{\mathrm{min}}$ $<$ $\epsilon$ $<$ $E_{\mathrm{max}}$, 
DOS $D(\epsilon)$ has three singularity points, i.e., the VHSs. The maximum, saddle, and minimum-point-type VHSs are located 
at $\epsilon$ = $E_{\mathrm{max}}$, $E_s$, and $E_{\mathrm{min}}$, respectively. 
As our calculated system was 2D, $D(\epsilon)$ was similar to a step function 
at $\epsilon$ = $E_{\mathrm{min}}$ and $E_{\mathrm{max}}$. 
At $\epsilon$ = $E_s$, $D(\epsilon)$ diverged 
\cite{Van_Hove_Phys_Rev_89_1189_1953, book_Grosso_2014}. 
%

\subsection{Transport Properties}

Figure \ref{fig_kagome_lattice_sigma_alpha}(a) shows the chemical potential $\mu$ dependence 
(carrier-concentration dependence) of 
the electrical conductivity $\sigma_{xx}$ at 0 K. 
In this case, we approximated the relaxation time $\tau$ as a constant of 10 fs. 
Additionally, $\sigma_{xx}$ is calculated using Eq. \eqref{eq_sigmaxx}. 
A large $\sigma_{xx}$ value was obtained at $\mu$ = $-0.15$ eV. 
From Eq. \eqref{eq_sigma_xx_dos_v} in Appendix \ref{section_Berry_energy}, a large $\sigma_{xx}$ value is obtained when 
the product of $D(\mu)$ and $v_x^2 (\mu)$ are large. 
Figure \ref{fig_kagome_lattice_band_dos}(c) shows that $D(\mu)$ peaks at $\mu$ = $E_s$ = $-0.18$ eV. 
The difference between the position of the DOS peak 
and that of the $\sigma_{xx}$ maximum was attributed to the group velocity $v_x$. 

Figure  \ref{fig_kagome_lattice_sigma_alpha}(b) shows the chemical potential $\mu$ dependence of 
the anomalous Hall conductivity $\sigma_{xy}$ at 0 K. 
In this study, $\sigma_{xy}$ is calculated using Eq. \eqref{eq_sigmaxy_berry}, 
considering the case in which the intrinsic contribution is dominant. 
Hereafter, we consider this case. 
In this case, $\sigma_{xy}$ = $-1$ $(e^2/h)$ for $\mu$ $<$ $E_{\mathrm{min}}$, 
$\sigma_{xy}$ = $-0.5$ $(e^2/h)$ for $\mu$ = $E_s$, and $\sigma_{xy}$ = 0 $(e^2/h)$ for $\mu$ $>$ $E_{\mathrm{max}}$. 
When one hole was doped, the quantized anomalous Hall conductivity $\sigma_{xy}$ = $-1$ ($e^2/h$) was obtained. 
In addition, the Chern numbers for the bottom, middle, and top bands were $-1$, 0, and 1, respectively. 
This result is consistent with the model calculation reported  by Ohgushi $et$ $al$. \cite{Ohgushi_Phys_Rev_B_62_R6065_2000}. 

To clarify the origin of $\sigma_{xy}$, we calculated the Berry curvature of the top band of the valence bands. 
Figure \ref{fig_kagome_lattice_berry}(a) shows the Berry curvature $\Omega_z(\boldsymbol{k})$ in $k$-space. 
Figure \ref{fig_kagome_lattice_berry}(b) shows the chemical-potential-dependent Berry curvature $\Omega_z(\mu)$ 
(refer to Eq. \eqref{eq_Berry_curvature_energy} in Appendix \ref{section_Berry_energy}). 
From Fig.  \ref{fig_kagome_lattice_berry}(a), 
$\Omega_z (\boldsymbol{k})$ = 0 at the $\Gamma$ point, and $\Omega_z(\boldsymbol{k})$ is maximum at the K point. 
This is because the energy difference between the top and middle bands is large at the $\Gamma$ point, 
whereas the energy difference is small at the K point, as shown 
in Fig. \ref{fig_kagome_lattice_band_dos}(a). 
From the relation between the energy difference and the magnitude of Berry curvature 
in Eq. \eqref{eq_def_berry} in Appendix \ref{sec_berry_curvature_chern_number}, 
$\Omega_z(\boldsymbol{k})$ was expected to be small 
at the $\Gamma$ point and large at the K point. 
Figure \ref{fig_kagome_lattice_berry}(a) shows that $\Omega_z (\boldsymbol{k})$ = 0 at the $\Gamma$ point 
and $\Omega_z (\boldsymbol{k})$ is maximum at the K point. 
Similarly, Fig. \ref{fig_kagome_lattice_berry}(b) shows that $\Omega_z (\mu)$ = 0 at $\mu$ = $E_{\mathrm{max}}$ 
and $\Omega_z (\mu)$ is maximum at $\mu$ = $E_{\mathrm{min}}$.  
From $\mu$ = $E_{\mathrm{min}}$ to $E_{\mathrm{max}}$, where $\Omega_z(\mu)$ was large, 
$\sigma_{xy}$ increased from $-1$ ($e^2/h$) to 0 ($e^2/h$). For $\mu$ $<$ $E_{\mathrm{min}}$ 
and $\mu$ $>$ $E_{\mathrm{max}}$, where $\Omega_z(\mu)$ = 0 , $\sigma_{xy}$ was constant. 
This is consistent with Eq. \eqref{eq_sigma_xy_dos_berry} in Appendix \ref{section_Berry_energy}, 
where $\sigma_{xy}$ is expressed in terms of $D(\mu)$ and $\Omega_z(\mu)$. 

\subsection{ Thermoelectric Conductivity $\alpha_{ij}$ }

Figure \ref{fig_kagome_lattice_sigma_alpha}(c) shows the chemical potential $\mu$ dependence of 
the longitudinal thermoelectric conductivity $\alpha_{xx}$. 
Large $|\alpha_{xx}|$ values were obtained at $\mu$ = $E_{\mathrm{min}}$, $E_s$, and $E_{\mathrm{max}}$. 
According to  Eq. \eqref{eq_alpha_xx}, $\alpha_{xx}$ is proportional to 
$\frac{d}{d\mu} ( D(\mu) v_x^2(\mu)) = \frac{dD}{d\mu} v_x^2 + D \frac{d v_x^2}{d\mu}$. 
As mentioned in Sect. \ref{section_introduction},  
at the saddle-point-type VHS, the DOS diverges in the 2D system. 
Therefore, at $\mu$ = $E_s$, $|D|$ and $| dD/d\mu |$ were large, resulting in the largest $|\alpha_{xx}|$ at 10 K. 
In contrast, for the maximum and minimum-point-type VHSs, 
the DOSs were similar to a step function. Therefore, $| dD/d\mu |$ was large at $\mu$ = $E_{\mathrm{min}}$ and $E_{\mathrm{max}}$.
Therefore, the $|\alpha_{xx}|$ values were large at the VHSs.

We obtained large $|\alpha_{xx}|$ values at the VHSs when the relaxation time $\tau$ was approximated as a constant. 
In the experimental studies on Bi$_2$Sr$_2$Ca$_{1-x}$Y$_x$Cu$_2$O$_{8+y}$ 
\cite{Munakata_Phys_Rev_B_45_10604_1992} 
and single-walled carbon nanotubes \cite{Yanagi_Nano_Lett_14_6437_2014}, 
the VHSs near the Fermi level were reported to enhance the Seebeck coefficients $|S|$. 
However, in a theoretical study on cuprate superconductors, using the energy $\epsilon$ dependent relaxation time $\tau (\epsilon)$, 
$|S|$ = 0 when the Fermi level was at the VHSs 
\cite{Newns_Phys_Rev_Lett_73_1695_1994}. 
These results indicate that the magnitude of $|\alpha_{xx}|$ may depend on the energy-dependent relaxation time $\tau (\epsilon)$ near the VHSs. 
Additionally, Eq. \eqref{eq_sigmaxy_berry} reveals that the anomalous Hall conductivity $\sigma_{xy}$ 
originating from the Berry curvature is independent of $\tau$. 
Therefore, there is no uncertainty in $\tau$ with respect to $\alpha_{xy}$. 

Figure \ref{fig_kagome_lattice_sigma_alpha}(d) shows the chemical potential $\mu$ dependence of $\alpha_{xy}$. 
The large $|\alpha_{xy}|$ values were obtained at $\mu$ = $E_{\mathrm{min}}$ and $E_s$, 
where the VHSs and the Berry curvatures are large. 
First, we considered $|\alpha_{xy}|$ at the saddle-point-type VHS ($\mu$ = $E_s$). 
From Eq. \eqref{eq_alpha_dos_berry}, 
at low temperatures, $\alpha_{xy}$ is proportional to the product of $D(\mu)$ and $\Omega_z (\mu)$. 
As $D(\mu)$ diverged at $\mu$ = $E_s$, the largest $|\alpha_{xy}|$ was obtained at $\mu$ = $E_s$. 
In addition, at $\mu$ = $E_{\mathrm{min}}$, $\Omega_z (\mu)$ was the largest.  
Therefore, a large $|\alpha_{xy}|$ was obtained even at $\mu$ = $E_{\mathrm{min}}$. 
In contrast, $\alpha_{xy}$ = 0 at $\mu$ = $E_{\mathrm{max}}$. 
This is because $\Omega_z (\mu)$ = 0 at $\mu$ = $E_{\mathrm{max}}$, 
as $\Omega_z (\boldsymbol{k})$ = 0 at the $\Gamma$ point. 
Therefore, the product of $D(\mu)$ and $\Omega_z (\mu)$ was 0, resulting in $\alpha_{xy}$ = 0.   
Thus, the large $\alpha_{xy}$ can be attributed to the VHSs and large Berry curvatures. 
More specifically, $|\alpha_{xy}|$ was large at $\mu$ = $E_s$ and $E_{\mathrm{min}}$ whereas 
$|\alpha_{xy}|$=0 at $\mu$ = $E_{\mathrm{max}}$. 

Large $|\alpha_{xy}|$ were obtained at $\mu$ = $E_{\mathrm{min}}$ and $E_s$ in our computed system. 
We examined the condition of the DOS and Berry curvatures that yielded large $|\alpha_{xy}|$. 
According to Eq. \eqref{eq_alpha_dos_berry}, both $D(\mu)$ and $\Omega_z (\mu)$ are important for the enhancement of $\alpha_{xy}$, 
as $\alpha_{xy}$ is proportional to the product $D(\mu) \Omega_z(\mu)$. 
To obtain large $|\alpha_{xy}|$ in the range of 1.0-10 ($\mu$V/K) ($e^2/h$) at 10 K, 
the product $D(\mu) \Omega_z(\mu)$ must be in the range of 30-300 eV$^{-1}$ \AA$^2$, 
as $\pi^2 e k_B^2 / 3 \hbar V$ was approximately 4.07 $\times$ 10$^{-3}$ ($\mu$V/K) ($e^2/h$) (eV/\AA$^2$) (1/K) in our case. 
Figure \ref{fig_kagome_lattice_sigma_alpha}(d) shows that large $|\alpha_{xy}|$ is obtained at $\mu$ = $E_{\mathrm{min}}$ and $E_s$. 
For $\mu$ = $E_{\mathrm{min}}$, the product $D(\mu) \Omega_z(\mu)$ was 97.9 eV$^{-1}$ \AA$^2$, 
as $D(E_{\mathrm{min}})$ was 5.47 eV$^{-1}$ and $\Omega_z(E_{\mathrm{min}})$ was 17.9 \AA$^2$. 
For $\mu$ = $E_s$, $D(\mu) \Omega_z(\mu)$ was 189 eV$^{-1}$ \AA$^2$, 
as $D(E_s)$ was 18.6 eV$^{-1}$ and $\Omega_z(E_s)$ was 10.2 \AA$^2$. 
Although $\Omega_z(\mu)$ at $\mu$ = $E_s$ was approximately half the maximum in this band, $D(\mu)$ was the maximum resulting in large $|\alpha_{xy}|$. 
By contrast, at $\mu$ = $E_{\mathrm{min}}$, $D(\mu)$ was less than 1/3 of $E_s$. 
As $\Omega_z(\mu)$ was maximum at $E_{\mathrm{min}}$, the product $D(\mu) \Omega_z(\mu)$ was of the same order as that of $E_s$. 
However, near $\mu$ = $E_{\mathrm{max}}$, $D(\mu) \Omega_z(\mu)$ was 0 because $\Omega_z(\mu)$ = 0. 
Thus, the large $|\alpha_{xy}|$ value originated from $D(\mu)$ was at $\mu$ = $E_s$, 
whereas the magnitudes of $\alpha_{xy}$ originated from $\Omega_z(\mu)$ were at $\mu$ = $E_{\mathrm{min}}$ and $E_{\mathrm{max}}$. 

Finally, we discussed the pure Nernst coefficient $N_0$ = $\alpha_{xy} / \sigma_{xx}$. 
From Appendix \ref{section_relaxation_time}, for the saddle-point-type VHS ($\mu$ = $E_s$), 
the intrinsic contribution in $\sigma_{xy}$ was dominant when the relaxation time $\tau$ was in the range of 10-100 fs. 
Table \ref{table_N} lists the values of $\sigma_{xy}$, $\tau$, $\sigma_{xx}$, $\theta_H$, and $N_0$ for $\mu$ = $E_s$. 
Additionally, 
$\tau$ was assumed to be a constant of 10 or 100 fs. 
For $\tau$ = 10 fs and $T$ = 50 K, a large value of $|N_0|$ = 9.74 $\mu$V/K was obtained. 
In our calculated system, the bandwidth was 0.2 eV when the lattice constant was set to 6.6 \AA. 
The bandwidths of FeCl$_2$ \cite{Syariati_APL_Materials_8_041105_2020} 
and CrI$_3$ \cite{Zhu_Appl_Phys_Lett_116_022404_2020}, 
which are 2D systems reported as realistic systems, 
are comparable to the bandwidth reported in the present study. 
Therefore, large $|N_0|$ can be obtained in 2D magnetic materials. 
Notably, a value of $|N_0|$ $\simeq$ 6 $\mu$V/K at 50 K was obtained 
from the first-principles calculations for FeCl$_2$ \cite{Syariati_APL_Materials_8_041105_2020}. 
Furthermore, for the 3D systems Fe$_3$Ga and Fe$_3$Al, $|N|$ values of approximately 1 $\mu$V/K
were obtained experimentally at 50 K \cite{Sakai_Nature_581_53_2020}; 
our $|N_0|$ results were approximately 10 times larger 
than the aforementioned value.

\begin{table}[hbtp]
  \caption{Values at $\mu$ = $E_s$ = $-0.18$ eV}
  \label{table_N}
  \centering
  \begin{tabular}{cccccc}
    \hline
    $T$ [K]     & $\sigma_{xy}$ [$e^2/h$] & $\tau$ [fs] & $\sigma_{xx}$ [$e^2/h$]  & $\theta_H$ & $N_0$ [$\mu$V/K] \\
    \hline \hline
    10             & $-0.459$                          &  10           & 2.16                                 & $-0.213$     & $-2.45$ \\
    10             & $-0.459$                          & 100          & 21.6                                 & $-0.0213$   & $-0.245$ \\
    50             & $-0.474$                          & 10            & 2.07                                 & $-0.230$     & $-9.74$ \\
    50             & $-0.474$                          & 100          & 20.7                                 & $-0.0230$   & $-0.974$ \\
    \hline
  \end{tabular}
\end{table}

\section{Summary}

We performed first-principles calculations based on DFT for a kagome lattice model with a chiral spin state, 
which is a typical model for a 2D magnetic system with a sizeable thermoelectric coefficient. 
First, the electronic structure was calculated; the density of states (DOS) analysis confirmed that 
maximum, minimum, and saddle-point-type van Hove singularities (VHSs) 
were present. 
We investigated the contributions of the VHSs to the thermoelectric coefficients. Large $|\alpha_{xx}|$ were obtained for VHSs with constant relaxation time approximation. 
Large $|\alpha_{xy}|$ were obtained at the VHSs at which the Berry curvatures were large. We derived an expression for the relationship between $\alpha_{ij}$ and the DOS based on the Mott relation as shown in Eqs. \eqref{eq_alpha_xx} and \eqref{eq_alpha_dos_berry}.
This expression shows that $\alpha_{xx}$ is proportional the energy derivative of 
$D(\epsilon) v_x^2(\epsilon)$.
Similarly, $\alpha_{xy}$ is proportional to $D(\epsilon) \Omega_z (\epsilon)$. 
These expressions clearly explain $\alpha_{xx}$ and $\alpha_{xy}$ at the VHSs.

The saddle-point-type VHS is particularly essential for thermoelectric effects in 2D systems. 
The largest $|\alpha_{xy}|$ was obtained at the chemical potential of the saddle-point-type VHS. 
In our calculated system, large Nernst coefficients of 10 $\mu$V/K at 50 K can be expected. The divergence of the DOS at the saddle-point-type VHS in 2D systems is one of the origins of large $|S|$ and $|N|$. 
Furthermore, 2D magnetic materials such as 
VSe$_2$ \cite{Bonilla_Nat_Nanotechnol_13_289_2018, Yu_Adv_Mater_31_1903779_2019}, 
Cr$_3$Te$_4$ \cite{Chua_Adv_Mater_33_2103360_2021}, 
CrTe \cite{Wu_Nat_Commun_12_5688_2021}, 
MnSe$_x$ \cite{OHara_Nano_Lett_18_3125_2018}, 
Fe$_3$GaTe$_2$ \cite{Zhang_Nat_Commun_13_5067_2022}, 
$\varepsilon$-Fe$_2$O$_3$ \cite{Yuan_Nano_Lett_19_3777_2019}, 
and 
CoFe$_2$O$_4$ \cite{Cheng_Nat_Commun_13_5241_2022}  
are good candidates for thermoelectric devices.

\section*{Acknowledgments}
This work was supported by JSPS KAKENHI 
(Grant Numbers JP20K15115, JP22K04862, JP22H05452, JP22H01889, and JP23H01129), JST SPRING (Grant Number JPMJSP2135), 
and the JST SICORP Program (Grant Number JPMJSC21E3).
The computation reported in this work was conducted using the facilities of the Supercomputer Center, 
the Institute for Solid State Physics, the University of Tokyo.

\appendix
\renewcommand{\thetable}{\Alph{section}.\arabic{figure}}
\renewcommand{\thefigure}{\Alph{section}.\arabic{table}}

\section{Thermoelectric Conductivity Expressed in terms of the Density of States}
\label{section_Berry_energy}
The DOS $D(\epsilon)$ can be expressed in terms of the Dirac delta function $\delta$ 
using the $n$-th quantum state energy $\epsilon_n (\bm{k})$ as follows: 
\begin{eqnarray}
D(\epsilon) = \sum_n V \int \frac{d^3 k}{(2 \pi)^3} \delta (\epsilon - \epsilon_n (\bm{k})),  \nonumber
\end{eqnarray}
where 
$V$, $\epsilon$, and $k$ are the unit cell volume, energy, and wavenumber, respectively. 
The anomalous Hall conductivity $\sigma_{xy} (\mu, T)$ can be calculated from the Berry curvature $\Omega_{n, z} (\bm{k})$ 
and the Fermi-Dirac distribution function $f(\epsilon, \mu, T)$ as follows: 
\begin{align}
\sigma_{xy}(\mu, T) =  \frac{e^2}{\hbar} \sum_n  \int \frac{d^3 k}{(2 \pi)^3} ~ \Omega_{n,z} (\bm{k})  f(\epsilon_n (\bm{k}), \mu, T), 
\label{eq_sigmaxy_3d}
\end{align}
where $\mu$ is the chemical potential. 
Changing the integrating variable in Eq. \eqref{eq_sigmaxy_3d} from $\bm{k}$ to $\epsilon$ yields 
\begin{align}
&\sigma_{xy}(\mu, T) \nonumber \\
&=  \frac{e^2}{\hbar} \sum_n  \int \frac{d^3 k}{(2 \pi)^3} ~ \Omega_{n,z} (\bm{k}) \int d\epsilon ~ \delta (\epsilon - \epsilon_n (\bm{k}))  f(\epsilon, \mu, T) 
\nonumber \\
&=
\frac{e^2}{\hbar} \frac{1}{V} \! \int \! d\epsilon \! \left[ \sum_{n} V \!\! \int \!\! \frac{d^3k}{(2\pi)^3} \Omega_{n,z} (\bm{k}) \delta (\epsilon - \epsilon_n (\bm{k})) \right] \!
 f(\epsilon, \mu, T) 
\nonumber \\
&=  \frac{e^2}{\hbar} \frac{1}{V}  \int d \epsilon  ~ D(\epsilon)  \Omega_z (\epsilon)  f(\epsilon, \mu, T) , 
\label{eq_sigma_xy_dos_berry}
\end{align}
where $\Omega_{z} (\epsilon)$ is defined as  
\begin{align}
\Omega_{z} (\epsilon) \equiv 
\frac{\sum_{n} V \int \frac{d^3k}{(2\pi)^3} \Omega_{n,z} (\bm{k}) \delta (\epsilon - \epsilon_n (\bm{k}))}
{\sum_{n} V \int \frac{d^3k}{(2\pi)^3}  \delta (\epsilon - \epsilon_{n} (\bm{k}))} . 
\nonumber 
\end{align}
At $T=0$ K, 
\begin{align}
\sigma_{xy}(\mu, T=0) =  \frac{e^2}{\hbar} \frac{1}{V}  \int_{- \infty}^{\mu} d \epsilon  ~ D(\epsilon)  \Omega_z (\epsilon)   . \nonumber
\end{align}
Differentiating both sides by $\mu$ results in 
\begin{align}
\frac{d \sigma_{xy}(\mu, T=0) }{d \mu} 
= 
 \frac{e^2}{\hbar} \frac{1}{V} D(\mu) \Omega_z (\mu) . \nonumber
\end{align}
Thus, the Berry curvature $\Omega_z (\mu)$ can be expressed by $d \sigma_{xy} / d \mu$ as follows: 
\begin{align}
\Omega_z (\mu) 
= 
\frac{\hbar V}{e^2 D(\mu)}  \frac{d \sigma_{xy} (\mu, T=0)}{d \mu} . 
\label{eq_Berry_curvature_energy}
\end{align}

At low temperatures, the Mott relation  
\cite{Smrcka_J_Phys_C_1977} 
is a good approximation, resulting in 
\begin{align}
\alpha_{ij} (T,\mu) = -  \frac{\pi^2 k_B^2}{3 |e|} T \left[ \frac{d \sigma_{ij} (T=0, \epsilon)}{d \epsilon} \right]_{\epsilon = \mu}, 
\label{eq_Mott}
\end{align}
where $k_B$ is the Boltzmann constant. 
The Mott relation (Eq. \eqref{eq_Mott}) clearly indicates that $\alpha_{xy}/T$ is constant 
regardless of temperature. 
Some experimental results have shown that 
$\alpha_{xy}$ is proportional to $-T$ at low temperatures and 
$- T \ln{(T)}$ at high temperatures for 
Co$_2$MnGa \cite{Sakai_Nat_Phys_14_1119_2018}, 
Fe$_3$$X$ ($X$ = Ga, Al) \cite{Sakai_Nature_581_53_2020}, 
and CoMnSb \cite{Nakamura_Phys_Rev_B_104_L161114_2021}.

Using the Mott relation (Eq. \eqref{eq_Mott}), at low temperatures, 
$\alpha_{xy}(\mu, T)$ can be expressed by $D(\mu) \Omega (\mu)$ as follows: 
\begin{align}
\alpha_{xy} (\mu, T) = - \frac{\pi^2  |e| k_B^2 T}{3 \hbar V}  D(\mu) \Omega (\mu) . \nonumber
\end{align}

With respect to $\sigma_{xx}$, 
if the relaxation time $\tau$ is assumed to be constant,  
Eq. \eqref{eq_sigmaxx} can be expressed using $D(\mu)$ as follows: 
\begin{align}
\sigma_{xx} (\mu, T) 
= 
e^2 \tau \frac{1}{V} \int d \epsilon ~ D(\epsilon) v_x^2 (\epsilon) 
\left(- \frac{\partial f}{\partial \epsilon} \right), 
\label{eq_sigma_xx_dos_v}
\end{align}
where $v_x^2 (\epsilon)$ is defined as  
\begin{align}
v_x^2 (\epsilon) 
\equiv 
\frac{\sum_{n} V \int \frac{d^3k}{(2\pi)^3} v_{n,x}^2 (\bm{k}) \delta (\epsilon - \epsilon_n (\bm{k}))}
{\sum_{n} V \int \frac{d^3k}{(2\pi)^3}  \delta (\epsilon - \epsilon_{n} (\bm{k}))} . 
\nonumber
\end{align}
As $- (\partial f / \partial \epsilon ) = \delta (\epsilon - \mu)$ at $T$ = 0 K, 
\begin{align}
\sigma_{xx} (\mu, T=0) = e^2 \tau \frac{1}{V} D(\mu) v_x^2 (\mu) . \nonumber
\end{align}
Differentiating both sides by $\mu$ results in 
\begin{align}
\frac{d \sigma_{xx} (\mu, T=0)}{d \mu} = e^2 \tau \frac{1}{V} \frac{d}{d\mu} \Bigl( D(\mu) v_x^2 (\mu) \Bigr) . \nonumber
\end{align}
By using the Mott relation (Eq. \eqref{eq_Mott}), at low temperatures, 
$\alpha_{xx}(\mu, T)$ can be expressed using $D(\mu) v_x^2(\mu)$ as follows: 
\begin{align}
\alpha_{xx} (\mu, T) = - \frac{\pi^2 |e| k_B^2 T \tau}{3 V} \frac{d}{d\mu} \Bigl( D(\mu) v_x^2 (\mu) \Bigr)  . \nonumber
\end{align}

\section{Violation of the Mott Relation at the VHSs}
\label{section_Mott_relation_VHS}

\begin{figure*}[htbp]
 \begin{center}
  \includegraphics[width=\linewidth]{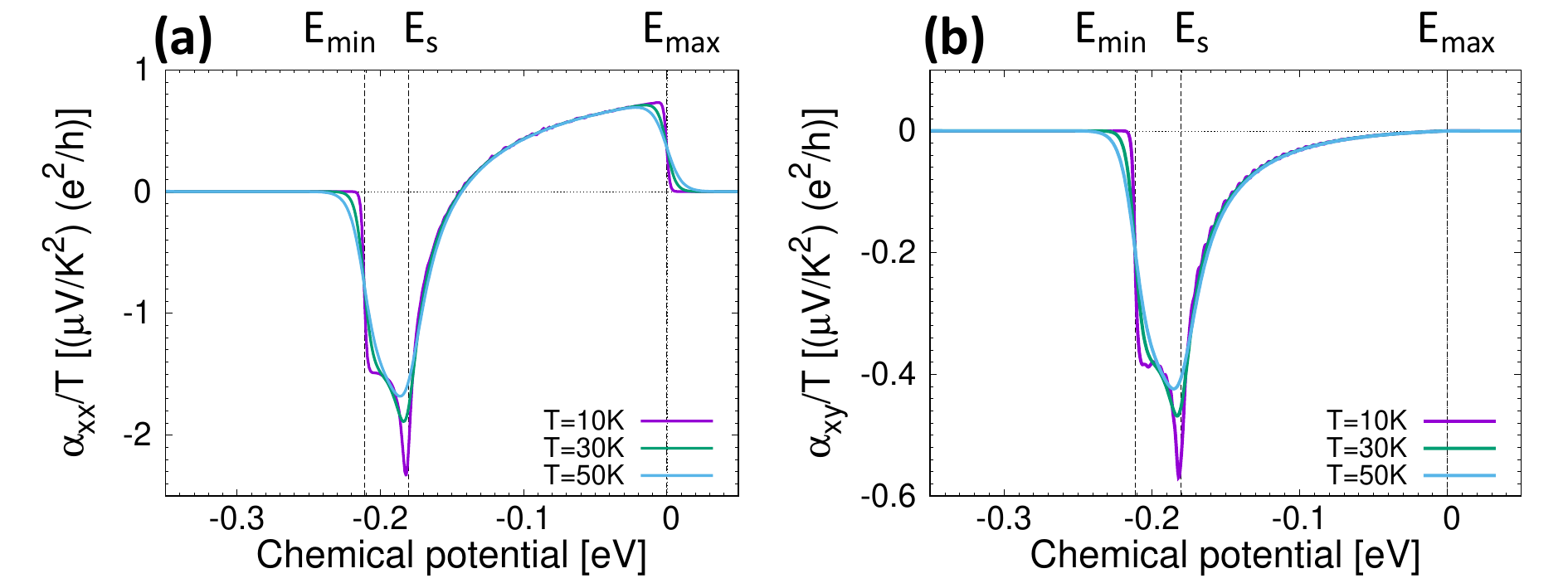}
 \end{center}
 \caption{(Color)
(a) Chemical potential dependence of $\alpha_{xx}/T$ 
at temperatures $T$ = 10, 30, and 50 K. 
Here, $\alpha_{xx}/T$ depends on the temperature near $E_{\mathrm{min}}$, $E_s$, and $E_{\mathrm{max}}$, where the VHSs were observed. 
The temperature dependence of $\alpha_{xx}/T$ is observed at $E_{\mathrm{min}}$, $E_s$, and $E_{\mathrm{max}}$ where the VHSs were observed. 
In other places, $\alpha_{xx}/T$ is constant regardless of temperature. 
(b) Chemical potential dependence of $\alpha_{xy}/T$. 
For temperatures $T$ = 10, 30, and 50 K, the values of $\alpha_{xx}/T$ depend on the temperature at $E_{\mathrm{min}}$ and $E_s$. 
In other places, $\alpha_{xy}/T$ is constant regardless of temperature. 
 }
 \label{fig_kagome_lattice_alpha_T}
\end{figure*}

From the Mott relation (Eq. \eqref{eq_Mott}), $\alpha_{xx}/T$ and $\alpha_{xy}/T$ should be constant regardless of temperature. 
However, Fig. \ref{fig_kagome_lattice_alpha_T}(a) shows that $\alpha_{xx}/T$ is dependent on the temperature 
at the VHSs ($\mu$ = $E_{\mathrm{min}}$, $E_s$, and $E_{\mathrm{max}}$), indicating the violation of the Mott relation. 
Figure \ref{fig_kagome_lattice_alpha_T}(b) shows that $\alpha_{xy}/T$ was temperature dependent 
near the minimum-point-type VHS ($E_{\mathrm{min}}$) and the saddle-point-type VHS ($E_s$).  
Minami $et$ $al$. 
\cite{Minami_Phys_Rev_B_102_205128_2020} 
reported that at the VHSs derived from nodal lines, $\alpha_{xy}/T$ is temperature-dependent.  
Although no nodal line exists in our system, we obtained similar results. 
Furthermore, $\alpha_{xx}/T$ was found to be temperature-dependent at the VHSs.

\section{Relaxation Time}
\label{section_relaxation_time}

Anomalous Hall conductivity $\sigma_{xy}$ has two types of origin contributions: extrinsic and intrinsic. 
The extrinsic contributions are the skew scattering 
\cite{
Smit_Physica_21_877_1955, 
Smit_Physica_24_39_1958} 
and the side jump 
\cite{
Berger_Phys_Rev_B_2_4559_1970, 
Berger_Phys_Rev_B_5_1862_1972}, 
whereas the intrinsic contribution is the Berry curvature 
\cite{
Xiao_Rev_Mod_Phys_82_1959_2010,  
Nagaosa_Rev_Mod_Phys_82_1539_2010}.
The anomalous Hall conductivity $\sigma_{xy}$ produced by the skew scattering is proportional 
to the electrical conductivity $\sigma_{xx}$. 
In contrast, $\sigma_{xy}$ produced by the side jump and Berry curvature is independent of $\sigma_{xx}$. 
Therefore, if the sample is clean and the temperature is low (large relaxation time), 
the skew scattering contribution is dominant in $\sigma_{xy}$ 
\cite{
Xiao_Rev_Mod_Phys_82_1959_2010, 
Nagaosa_Rev_Mod_Phys_82_1539_2010}. 
By contrast, if the sample is dirty and the temperature is high (small relaxation time), 
the contributions of the side jump and Berry curvature are dominant in $\sigma_{xy}$ 
\cite{
Xiao_Rev_Mod_Phys_82_1959_2010, 
Nagaosa_Rev_Mod_Phys_82_1539_2010}.  
Onoda $et$ $al$. 
\cite{Onoda_Phys_Rev_B_77_165103_2008} 
reported that in the moderately dirty regime with 
$\sigma_{xx}$ $\sim$ 3 $\times$ $10^3$ - 5 $\times$ $10^5$ $\Omega^{-1}$ cm$^{-1}$, 
$\sigma_{xy}$ is almost constant with values of the order of $10^2$-$10^3$ $\Omega^{-1}$ cm$^{-1}$. 
In this case, intrinsic contribution to the anomalous Hall conductivity $\sigma_{xy}$ is dominant. 

In our system, when the relaxation time $\tau$ was assumed to have constant values in the range 
$\tau$ = 10-100 fs, the intrinsic contribution dominated in the anomalous Hall conductivity $\sigma_{xy}$. 
This case is consistent with the real system. 
As our calculated system was a monolayer, the thickness was considered to be 1 \AA. 
In this case, at $\mu$ = $E_s$, $\sigma_{xy}$ $\simeq$ $-2$ $\times$ 10$^3$ $\Omega^{-1}$ cm$^{-1}$. 
We found that $\sigma_{xx}$ $\simeq$ 1 $\times$ 10$^4$ $\Omega^{-1}$ cm$^{-1}$ for $\tau$ = 10 fs, 
with the Hall angle $\theta_H$ = $\sigma_{xy} / \sigma_{xx}$ $\simeq$ $-0.2$. 
Moreover, $\sigma_{xx}$ $\simeq$ 1 $\times$ 10$^5$ $\Omega^{-1}$ cm$^{-1}$ was obtained for $\tau$ =100 fs, 
where $\theta_H$ $\simeq$ $-0.02$. 
Therefore, for the present system, a relaxation time in the range of 10-100 fs was consistent with a realistic system 
in the moderately dirty regime in which the intrinsic contribution was dominant.

\section{Computational Model}
\label{sec_computational_model}

\begin{figure}[htbp]
 \begin{center}
  \includegraphics[width=\linewidth]{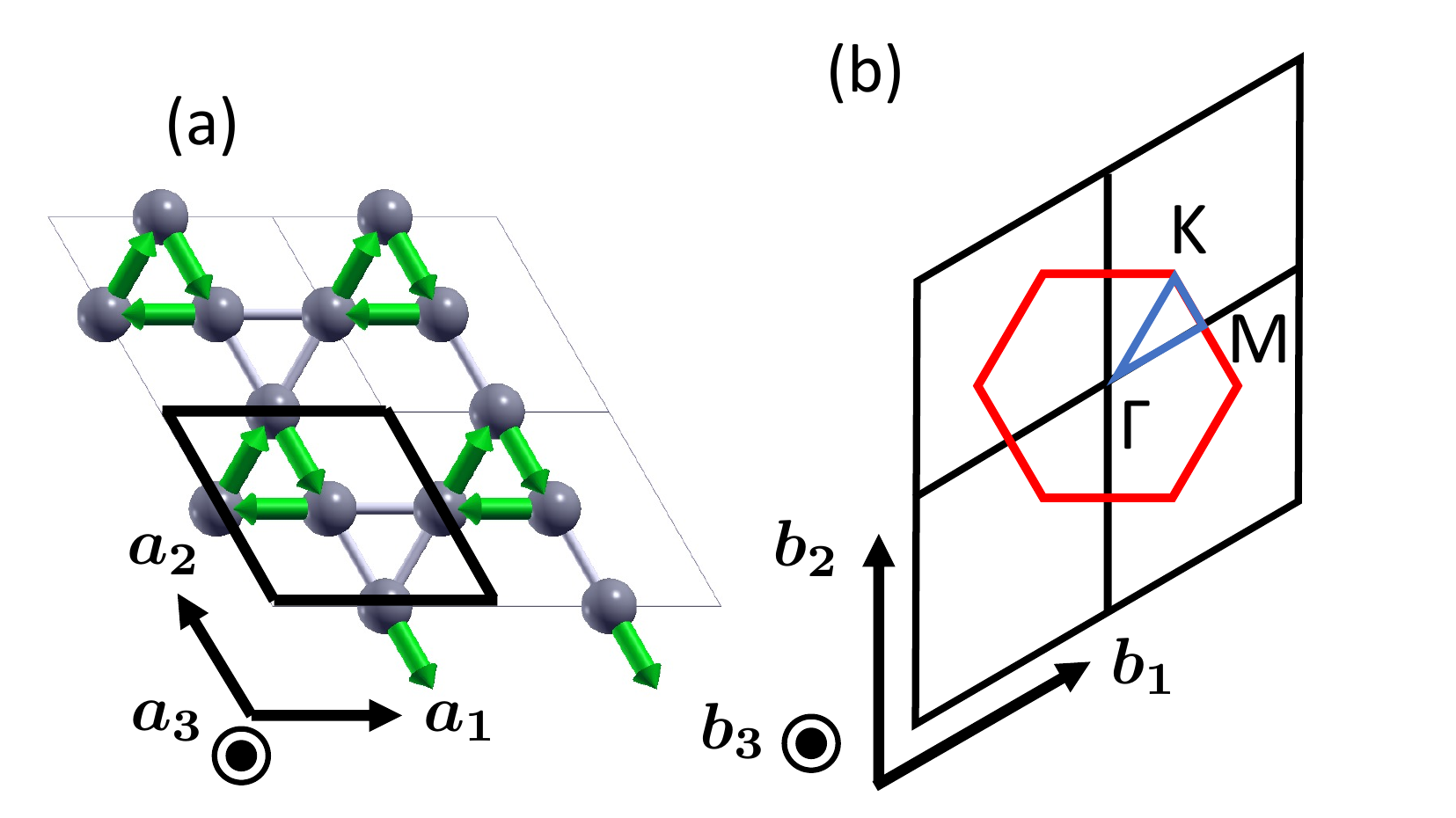}
 \end{center}
 \caption{(Color)
 (a) Top view of the computed model of the kagome lattice. 
 The black rhombus represents the unit cell with  
 $a_1$ = $a_2$ = 6.60 \AA, $a_3$ = 100 \AA,  
 $\alpha$ = 120$^\circ$, $\beta$ = 90$^\circ$, and $\gamma$ = 90$^\circ$. 
 The gray spheres represent fixed hydrogen atoms, and the gray lines represent chemical bonds. 
 The green arrows show the spin moment, where 
 $\bm{s}_1 = (\theta = 70^\circ, \phi = 60^\circ)$, 
 $\bm{s}_2 = (\theta = 70^\circ, \phi = 180^\circ)$, 
 and $\bm{s}_3 = (\theta = 70^\circ, \phi = 300^\circ)$. 
 (b) The reciprocal lattice of the kagome lattice, where  
 $\bm{b}_1$, $\bm{b}_2$, and $\bm{b}_3$ are the reciprocal lattice vectors.
 The red hexagon represents the first Brillouin zone. 
 The fractional coordinates of the reciprocal lattice space for the $\Gamma$, M, and K points 
 are (0, 0, 0), (1/2, 0, 0), and (1/3, 1/3, 0), respectively. 
 At these coordinates, $2\pi/a$ is set to 1, where $a$ is the length of the unit cell vector.
 }
 \label{fig_kagome_lattice_structure}
\end{figure}

Figure \ref{fig_kagome_lattice_structure}(a) is the computed model of the kagome lattice, 
and Fig. \ref{fig_kagome_lattice_structure}(b) is the reciprocal lattice. 

\section{Berry Curvature and Chern Number}
\label{sec_berry_curvature_chern_number}

The components of the Berry curvature $\Omega_{n, \mu \nu}  (\bm{k}) $ 
can be written as a summation over the periodic part of Bloch states $| u_{n} (\bm{k})  \rangle$ as follows 
\cite{
Thouless_Phys_Rev_Lett_49_405_1982,
Yao_Phys_Rev_Lett_92_037204_2004}: 
\begin{align}
\Omega_{n, \mu \nu} (\bm{k}) = -2 \hbar^2 ~ \mathrm{Im} \sum_{n' \ne n} 
\frac{ \langle u_{n} | v_\mu | u_{n'} \rangle  
\langle u_{n'} | v_\nu | u_{n} \rangle }
{( \epsilon_n - \epsilon_{n'} )^2} . 
\label{eq_def_berry}
\end{align} 
If all bands are separated at $T = 0$ K, the anomalous Hall conductivity $\sigma_{xy}$ 
is quantized by the Chern number $\nu_n$ in the unit of $e^2/h$ 
\cite{Klitzing_Phys_Rev_Lett_45_494_1980, 
Thouless_Phys_Rev_Lett_49_405_1982}.  
Furthermore, 
$\nu_n$ is calculated by integrating the Berry curvature $\Omega_{n,z}$ 
as follows: $\nu_n = \int \frac{d^2 k}{2\pi} ~ \Omega_{n, z} (\bm{k})$ 
\cite{
Xiao_Rev_Mod_Phys_82_1959_2010, 
Nagaosa_Rev_Mod_Phys_82_1539_2010}, 
and $\nu_n$ is always an integer. 
Therefore, the quantized anomalous Hall conductivity $\sigma_{xy}$ at $T = 0$ K 
is expressed as the sum of $\nu_n$ as follows: 
\begin{align}
\sigma_{xy} (T=0) =  \frac{e^2}{h} \sum_n \nu_n.
\label{eq_sigmaxy_chern}
\end{align}


\bibliography{paper_kagome_lattice_shibata_ref} 

\begin{thebibliography}{53}%
\makeatletter
\providecommand \@ifxundefined [1]{%
 \@ifx{#1\undefined}
}%
\providecommand \@ifnum [1]{%
 \ifnum #1\expandafter \@firstoftwo
 \else \expandafter \@secondoftwo
 \fi
}%
\providecommand \@ifx [1]{%
 \ifx #1\expandafter \@firstoftwo
 \else \expandafter \@secondoftwo
 \fi
}%
\providecommand \natexlab [1]{#1}%
\providecommand \enquote  [1]{``#1''}%
\providecommand \bibnamefont  [1]{#1}%
\providecommand \bibfnamefont [1]{#1}%
\providecommand \citenamefont [1]{#1}%
\providecommand \href@noop [0]{\@secondoftwo}%
\providecommand \href [0]{\begingroup \@sanitize@url \@href}%
\providecommand \@href[1]{\@@startlink{#1}\@@href}%
\providecommand \@@href[1]{\endgroup#1\@@endlink}%
\providecommand \@sanitize@url [0]{\catcode `\\12\catcode `\$12\catcode
  `\&12\catcode `\#12\catcode `\^12\catcode `\_12\catcode `\%12\relax}%
\providecommand \@@startlink[1]{}%
\providecommand \@@endlink[0]{}%
\providecommand \url  [0]{\begingroup\@sanitize@url \@url }%
\providecommand \@url [1]{\endgroup\@href {#1}{\urlprefix }}%
\providecommand \urlprefix  [0]{URL }%
\providecommand \Eprint [0]{\href }%
\providecommand \doibase [0]{https://doi.org/}%
\providecommand \selectlanguage [0]{\@gobble}%
\providecommand \bibinfo  [0]{\@secondoftwo}%
\providecommand \bibfield  [0]{\@secondoftwo}%
\providecommand \translation [1]{[#1]}%
\providecommand \BibitemOpen [0]{}%
\providecommand \bibitemStop [0]{}%
\providecommand \bibitemNoStop [0]{.\EOS\space}%
\providecommand \EOS [0]{\spacefactor3000\relax}%
\providecommand \BibitemShut  [1]{\csname bibitem#1\endcsname}%
\let\auto@bib@innerbib\@empty
\bibitem [{\citenamefont {Mizuta}\ and\ \citenamefont
  {Ishii}(2015)}]{Mizuta_JPS_Conf_Proc_5_011023_2015}%
  \BibitemOpen
  \bibfield  {author} {\bibinfo {author} {\bibfnamefont {Y.~P.}\ \bibnamefont
  {Mizuta}}\ and\ \bibinfo {author} {\bibfnamefont {F.}~\bibnamefont {Ishii}},\
  }\bibfield  {title} {\bibinfo {title} {{Thermopower of Doped Quantum
  Anomalous Hall Insulators: The case of Dirac Hamiltonian}},\ }\href
  {https://journals.jps.jp/doi/abs/10.7566/JPSCP.5.011023} {\bibfield
  {journal} {\bibinfo  {journal} {JPS Conf. Proc.}\ }\textbf {\bibinfo {volume}
  {5}},\ \bibinfo {pages} {011023} (\bibinfo {year} {2015})}\BibitemShut
  {NoStop}%
\bibitem [{\citenamefont {Mizuta}\ and\ \citenamefont
  {Ishii}(2016)}]{Mizuta_Sci_Rep_6_28076_2016}%
  \BibitemOpen
  \bibfield  {author} {\bibinfo {author} {\bibfnamefont {Y.~P.}\ \bibnamefont
  {Mizuta}}\ and\ \bibinfo {author} {\bibfnamefont {F.}~\bibnamefont {Ishii}},\
  }\bibfield  {title} {\bibinfo {title} {{Large anomalous Nernst effect in a
  skyrmion crystal}},\ }\href {https://doi.org/10.1038/srep28076} {\bibfield
  {journal} {\bibinfo  {journal} {Sci. Rep.}\ }\textbf {\bibinfo {volume}
  {6}},\ \bibinfo {pages} {28076} (\bibinfo {year} {2016})}\BibitemShut
  {NoStop}%
\bibitem [{\citenamefont {Mahan}\ and\ \citenamefont
  {Sofo}(1996)}]{Mahan_Proc_Natl_Acad_Sci_USA_93_7439_1996}%
  \BibitemOpen
  \bibfield  {author} {\bibinfo {author} {\bibfnamefont {G.~D.}\ \bibnamefont
  {Mahan}}\ and\ \bibinfo {author} {\bibfnamefont {J.~O.}\ \bibnamefont
  {Sofo}},\ }\bibfield  {title} {\bibinfo {title} {{The best thermoelectric}},\
  }\href {https://doi.org/10.1073/pnas.93.15.7436} {\bibfield  {journal}
  {\bibinfo  {journal} {Proc. Natl. Acad. Sci. USA}\ }\textbf {\bibinfo
  {volume} {93}},\ \bibinfo {pages} {7436} (\bibinfo {year}
  {1996})}\BibitemShut {NoStop}%
\bibitem [{\citenamefont {Minami}\ \emph {et~al.}(2020)\citenamefont {Minami},
  \citenamefont {Ishii}, \citenamefont {Hirayama}, \citenamefont {Nomoto},
  \citenamefont {Koretsune},\ and\ \citenamefont
  {Arita}}]{Minami_Phys_Rev_B_102_205128_2020}%
  \BibitemOpen
  \bibfield  {author} {\bibinfo {author} {\bibfnamefont {S.}~\bibnamefont
  {Minami}}, \bibinfo {author} {\bibfnamefont {F.}~\bibnamefont {Ishii}},
  \bibinfo {author} {\bibfnamefont {M.}~\bibnamefont {Hirayama}}, \bibinfo
  {author} {\bibfnamefont {T.}~\bibnamefont {Nomoto}}, \bibinfo {author}
  {\bibfnamefont {T.}~\bibnamefont {Koretsune}},\ and\ \bibinfo {author}
  {\bibfnamefont {R.}~\bibnamefont {Arita}},\ }\bibfield  {title} {\bibinfo
  {title} {{Enhancement of the transverse thermoelectric conductivity
  originating from stationary points in nodal lines}},\ }\href
  {https://doi.org/10.1103/PhysRevB.102.205128} {\bibfield  {journal} {\bibinfo
   {journal} {Phys. Rev. B}\ }\textbf {\bibinfo {volume} {102}},\ \bibinfo
  {pages} {205128} (\bibinfo {year} {2020})}\BibitemShut {NoStop}%
\bibitem [{\citenamefont {Van~Hove}(1953)}]{Van_Hove_Phys_Rev_89_1189_1953}%
  \BibitemOpen
  \bibfield  {author} {\bibinfo {author} {\bibfnamefont {L.}~\bibnamefont
  {Van~Hove}},\ }\bibfield  {title} {\bibinfo {title} {{The Occurrence of
  Singularities in the Elastic Frequency Distribution of a Crystal}},\ }\href
  {https://doi.org/10.1103/PhysRev.89.1189} {\bibfield  {journal} {\bibinfo
  {journal} {Phys. Rev.}\ }\textbf {\bibinfo {volume} {89}},\ \bibinfo {pages}
  {1189} (\bibinfo {year} {1953})}\BibitemShut {NoStop}%
\bibitem [{\citenamefont {Grosso}\ and\ \citenamefont
  {Parravicini}(2014)}]{book_Grosso_2014}%
  \BibitemOpen
  \bibfield  {author} {\bibinfo {author} {\bibfnamefont {G.}~\bibnamefont
  {Grosso}}\ and\ \bibinfo {author} {\bibfnamefont {G.~P.}\ \bibnamefont
  {Parravicini}},\ }\href@noop {} {\emph {\bibinfo {title} {Solid State Physics
  (Second Edition)}}}\ (\bibinfo  {publisher} {Academic Press Inc},\ \bibinfo
  {year} {2014})\BibitemShut {NoStop}%
\bibitem [{\citenamefont {Verzola}\ \emph {et~al.}(2022)\citenamefont
  {Verzola}, \citenamefont {Villaos}, \citenamefont {Purwitasari},
  \citenamefont {Huang}, \citenamefont {Hsu}, \citenamefont {Chang},
  \citenamefont {Lin},\ and\ \citenamefont
  {Chuang}}]{Verzola_Mater_Today_Commun_33_104468_2022}%
  \BibitemOpen
  \bibfield  {author} {\bibinfo {author} {\bibfnamefont {I.~M.~R.}\
  \bibnamefont {Verzola}}, \bibinfo {author} {\bibfnamefont {R.~A.~B.}\
  \bibnamefont {Villaos}}, \bibinfo {author} {\bibfnamefont {W.}~\bibnamefont
  {Purwitasari}}, \bibinfo {author} {\bibfnamefont {Z.-Q.}\ \bibnamefont
  {Huang}}, \bibinfo {author} {\bibfnamefont {C.-H.}\ \bibnamefont {Hsu}},
  \bibinfo {author} {\bibfnamefont {G.}~\bibnamefont {Chang}}, \bibinfo
  {author} {\bibfnamefont {H.}~\bibnamefont {Lin}},\ and\ \bibinfo {author}
  {\bibfnamefont {F.-C.}\ \bibnamefont {Chuang}},\ }\bibfield  {title}
  {\bibinfo {title} {{Prediction of van Hove singularities, excellent
  thermoelectric performance, and non-trivial topology in monolayer rhenium
  dichalcogenides}},\ }\href
  {https://doi.org/https://doi.org/10.1016/j.mtcomm.2022.104468} {\bibfield
  {journal} {\bibinfo  {journal} {Mater. Today Commun.}\ }\textbf {\bibinfo
  {volume} {33}},\ \bibinfo {pages} {104468} (\bibinfo {year}
  {2022})}\BibitemShut {NoStop}%
\bibitem [{\citenamefont {Syariati}\ \emph {et~al.}(2020)\citenamefont
  {Syariati}, \citenamefont {Minami}, \citenamefont {Sawahata},\ and\
  \citenamefont {Ishii}}]{Syariati_APL_Materials_8_041105_2020}%
  \BibitemOpen
  \bibfield  {author} {\bibinfo {author} {\bibfnamefont {R.}~\bibnamefont
  {Syariati}}, \bibinfo {author} {\bibfnamefont {S.}~\bibnamefont {Minami}},
  \bibinfo {author} {\bibfnamefont {H.}~\bibnamefont {Sawahata}},\ and\
  \bibinfo {author} {\bibfnamefont {F.}~\bibnamefont {Ishii}},\ }\bibfield
  {title} {\bibinfo {title} {{First-principles study of anomalous Nernst effect
  in half-metallic iron dichloride monolayer}},\ }\href
  {https://aip.scitation.org/doi/10.1063/1.5143474} {\bibfield  {journal}
  {\bibinfo  {journal} {APL Materials}\ }\textbf {\bibinfo {volume} {8}},\
  \bibinfo {pages} {041105} (\bibinfo {year} {2020})}\BibitemShut {NoStop}%
\bibitem [{\citenamefont {Xu}\ \emph {et~al.}(2019)\citenamefont {Xu},
  \citenamefont {Phelan},\ and\ \citenamefont
  {Chien}}]{Xu_Nano_Lett_19_8250_2019}%
  \BibitemOpen
  \bibfield  {author} {\bibinfo {author} {\bibfnamefont {J.}~\bibnamefont
  {Xu}}, \bibinfo {author} {\bibfnamefont {W.~A.}\ \bibnamefont {Phelan}},\
  and\ \bibinfo {author} {\bibfnamefont {C.-L.}\ \bibnamefont {Chien}},\
  }\bibfield  {title} {\bibinfo {title} {{Large Anomalous Nernst Effect in a
  van der Waals Ferromagnet Fe$_3$GeTe$_2$}},\ }\href
  {https://pubs.acs.org/doi/10.1021/acs.nanolett.9b03739} {\bibfield  {journal}
  {\bibinfo  {journal} {Nano Lett.}\ }\textbf {\bibinfo {volume} {19}},\
  \bibinfo {pages} {8250} (\bibinfo {year} {2019})}\BibitemShut {NoStop}%
\bibitem [{\citenamefont {Yang}\ \emph {et~al.}(2021)\citenamefont {Yang},
  \citenamefont {Zhou}, \citenamefont {Feng},\ and\ \citenamefont
  {Yao}}]{Yang_Phys_Rev_B_103_024436_2021}%
  \BibitemOpen
  \bibfield  {author} {\bibinfo {author} {\bibfnamefont {X.}~\bibnamefont
  {Yang}}, \bibinfo {author} {\bibfnamefont {X.}~\bibnamefont {Zhou}}, \bibinfo
  {author} {\bibfnamefont {W.}~\bibnamefont {Feng}},\ and\ \bibinfo {author}
  {\bibfnamefont {Y.}~\bibnamefont {Yao}},\ }\bibfield  {title} {\bibinfo
  {title} {{Tunable magneto-optical effect, anomalous Hall effect, and
  anomalous Nernst effect in the two-dimensional room-temperature ferromagnet
  $1T\ensuremath{-}{\mathrm{CrTe}}_{2}$}},\ }\href
  {https://doi.org/10.1103/PhysRevB.103.024436} {\bibfield  {journal} {\bibinfo
   {journal} {Phys. Rev. B}\ }\textbf {\bibinfo {volume} {103}},\ \bibinfo
  {pages} {024436} (\bibinfo {year} {2021})}\BibitemShut {NoStop}%
\bibitem [{\citenamefont {Minami}\ \emph {et~al.}(2018)\citenamefont {Minami},
  \citenamefont {Ishii}, \citenamefont {Mizuta},\ and\ \citenamefont
  {Saito}}]{Minami_Appl_Phys_Lett_113_032403_2018}%
  \BibitemOpen
  \bibfield  {author} {\bibinfo {author} {\bibfnamefont {S.}~\bibnamefont
  {Minami}}, \bibinfo {author} {\bibfnamefont {F.}~\bibnamefont {Ishii}},
  \bibinfo {author} {\bibfnamefont {Y.~P.}\ \bibnamefont {Mizuta}},\ and\
  \bibinfo {author} {\bibfnamefont {M.}~\bibnamefont {Saito}},\ }\bibfield
  {title} {\bibinfo {title} {{First-principles study on thermoelectric
  properties of half-Heusler compounds Co$M$Sb ($M$ = Sc, Ti, V, Cr, and
  Mn)}},\ }\href
  {https://aip.scitation.org/doi/10.1063/1.5029907?mi=3h7ls6&af=R&ConceptID=43&SeriesKey=apl&target=topic}
  {\bibfield  {journal} {\bibinfo  {journal} {Appl. Phys. Lett.}\ }\textbf
  {\bibinfo {volume} {113}},\ \bibinfo {pages} {032403} (\bibinfo {year}
  {2018})}\BibitemShut {NoStop}%
\bibitem [{\citenamefont {Xiao}\ \emph {et~al.}(2010)\citenamefont {Xiao},
  \citenamefont {Chang},\ and\ \citenamefont
  {Niu}}]{Xiao_Rev_Mod_Phys_82_1959_2010}%
  \BibitemOpen
  \bibfield  {author} {\bibinfo {author} {\bibfnamefont {D.}~\bibnamefont
  {Xiao}}, \bibinfo {author} {\bibfnamefont {M.-C.}\ \bibnamefont {Chang}},\
  and\ \bibinfo {author} {\bibfnamefont {Q.}~\bibnamefont {Niu}},\ }\bibfield
  {title} {\bibinfo {title} {{Berry phase effects on electronic properties}},\
  }\href {https://doi.org/10.1103/RevModPhys.82.1959} {\bibfield  {journal}
  {\bibinfo  {journal} {Rev. Mod. Phys.}\ }\textbf {\bibinfo {volume} {82}},\
  \bibinfo {pages} {1959} (\bibinfo {year} {2010})}\BibitemShut {NoStop}%
\bibitem [{\citenamefont {Nagaosa}\ \emph {et~al.}(2010)\citenamefont
  {Nagaosa}, \citenamefont {Sinova}, \citenamefont {Onoda}, \citenamefont
  {MacDonald},\ and\ \citenamefont {Ong}}]{Nagaosa_Rev_Mod_Phys_82_1539_2010}%
  \BibitemOpen
  \bibfield  {author} {\bibinfo {author} {\bibfnamefont {N.}~\bibnamefont
  {Nagaosa}}, \bibinfo {author} {\bibfnamefont {J.}~\bibnamefont {Sinova}},
  \bibinfo {author} {\bibfnamefont {S.}~\bibnamefont {Onoda}}, \bibinfo
  {author} {\bibfnamefont {A.~H.}\ \bibnamefont {MacDonald}},\ and\ \bibinfo
  {author} {\bibfnamefont {N.~P.}\ \bibnamefont {Ong}},\ }\bibfield  {title}
  {\bibinfo {title} {{Anomalous Hall effect}},\ }\href
  {https://doi.org/10.1103/RevModPhys.82.1539} {\bibfield  {journal} {\bibinfo
  {journal} {Rev. Mod. Phys.}\ }\textbf {\bibinfo {volume} {82}},\ \bibinfo
  {pages} {1539} (\bibinfo {year} {2010})}\BibitemShut {NoStop}%
\bibitem [{\citenamefont {Ozaki}(2003)}]{Ozaki_Phys_Rev_B_67_155108_2003}%
  \BibitemOpen
  \bibfield  {author} {\bibinfo {author} {\bibfnamefont {T.}~\bibnamefont
  {Ozaki}},\ }\bibfield  {title} {\bibinfo {title} {{Variationally optimized
  atomic orbitals for large-scale electronic structures}},\ }\href
  {https://doi.org/10.1103/PhysRevB.67.155108} {\bibfield  {journal} {\bibinfo
  {journal} {Phys. Rev. B}\ }\textbf {\bibinfo {volume} {67}},\ \bibinfo
  {pages} {155108} (\bibinfo {year} {2003})}\BibitemShut {NoStop}%
\bibitem [{\citenamefont {Ozaki}\ and\ \citenamefont
  {Kino}(2004)}]{Ozaki_Phys_Rev_B_69_195113_2004}%
  \BibitemOpen
  \bibfield  {author} {\bibinfo {author} {\bibfnamefont {T.}~\bibnamefont
  {Ozaki}}\ and\ \bibinfo {author} {\bibfnamefont {H.}~\bibnamefont {Kino}},\
  }\bibfield  {title} {\bibinfo {title} {{Numerical atomic basis orbitals from
  H to Kr}},\ }\href {https://doi.org/10.1103/PhysRevB.69.195113} {\bibfield
  {journal} {\bibinfo  {journal} {Phys. Rev. B}\ }\textbf {\bibinfo {volume}
  {69}},\ \bibinfo {pages} {195113} (\bibinfo {year} {2004})}\BibitemShut
  {NoStop}%
\bibitem [{\citenamefont {Ozaki}\ and\ \citenamefont
  {Kino}(2005)}]{Ozaki_Phys_Rev_B_72_045121_2005}%
  \BibitemOpen
  \bibfield  {author} {\bibinfo {author} {\bibfnamefont {T.}~\bibnamefont
  {Ozaki}}\ and\ \bibinfo {author} {\bibfnamefont {H.}~\bibnamefont {Kino}},\
  }\bibfield  {title} {\bibinfo {title} {{Efficient projector expansion for the
  ab initio LCAO method}},\ }\href {https://doi.org/10.1103/PhysRevB.72.045121}
  {\bibfield  {journal} {\bibinfo  {journal} {Phys. Rev. B}\ }\textbf {\bibinfo
  {volume} {72}},\ \bibinfo {pages} {045121} (\bibinfo {year}
  {2005})}\BibitemShut {NoStop}%
\bibitem [{\citenamefont {Lejaeghere}\ \emph {et~al.}(2016)\citenamefont
  {Lejaeghere}, \citenamefont {Bihlmayer}, \citenamefont {Bj\"{o}rkman},
  \citenamefont {Blaha}, \citenamefont {Bl\"{u}gel}, \citenamefont {Blum},
  \citenamefont {Caliste}, \citenamefont {Castelli}, \citenamefont {Clark},
  \citenamefont {Dal~Corso}, \citenamefont {De~Gironcoli}, \citenamefont
  {Deutsch}, \citenamefont {Dewhurst}, \citenamefont {Di~Marco}, \citenamefont
  {Draxl}, \citenamefont {Du\l{}ak}, \citenamefont {Eriksson}, \citenamefont
  {Flores-Livas}, \citenamefont {Garrity}, \citenamefont {Genovese},
  \citenamefont {Giannozzi}, \citenamefont {Giantomassi}, \citenamefont
  {Goedecker}, \citenamefont {Gonze}, \citenamefont {Gr\r{a}n\"{a}s},
  \citenamefont {Gross}, \citenamefont {Gulans}, \citenamefont {Gygi},
  \citenamefont {Hamann}, \citenamefont {Hasnip}, \citenamefont {Holzwarth},
  \citenamefont {Iu\c{s}an}, \citenamefont {Jochym}, \citenamefont {Jollet},
  \citenamefont {Jones}, \citenamefont {Kresse}, \citenamefont {Koepernik},
  \citenamefont {K\"{u}\c{c}\"{u}kbenli}, \citenamefont {Kvashnin},
  \citenamefont {Locht}, \citenamefont {Lubeck}, \citenamefont {Marsman},
  \citenamefont {Marzari}, \citenamefont {Nitzsche}, \citenamefont
  {Nordstr\"{o}m}, \citenamefont {Ozaki}, \citenamefont {Paulatto},
  \citenamefont {Pickard}, \citenamefont {Poelmans}, \citenamefont {Probert},
  \citenamefont {Refson}, \citenamefont {Richter}, \citenamefont {Rignanese},
  \citenamefont {Saha}, \citenamefont {Scheffler}, \citenamefont {Schlipf},
  \citenamefont {Schwarz}, \citenamefont {Sharma}, \citenamefont {Tavazza},
  \citenamefont {Thunstr\"{o}m}, \citenamefont {Tkatchenko}, \citenamefont
  {Torrent}, \citenamefont {Vanderbilt}, \citenamefont {Van~Setten},
  \citenamefont {Van~Speybroeck}, \citenamefont {Wills}, \citenamefont {Yates},
  \citenamefont {Zhang},\ and\ \citenamefont
  {Cottenier}}]{Lejaeghere_Science_351_aad3000_2016}%
  \BibitemOpen
  \bibfield  {author} {\bibinfo {author} {\bibfnamefont {K.}~\bibnamefont
  {Lejaeghere}}, \bibinfo {author} {\bibfnamefont {G.}~\bibnamefont
  {Bihlmayer}}, \bibinfo {author} {\bibfnamefont {T.}~\bibnamefont
  {Bj\"{o}rkman}}, \bibinfo {author} {\bibfnamefont {P.}~\bibnamefont {Blaha}},
  \bibinfo {author} {\bibfnamefont {S.}~\bibnamefont {Bl\"{u}gel}}, \bibinfo
  {author} {\bibfnamefont {V.}~\bibnamefont {Blum}}, \bibinfo {author}
  {\bibfnamefont {D.}~\bibnamefont {Caliste}}, \bibinfo {author} {\bibfnamefont
  {I.~E.}\ \bibnamefont {Castelli}}, \bibinfo {author} {\bibfnamefont {S.~J.}\
  \bibnamefont {Clark}}, \bibinfo {author} {\bibfnamefont {A.}~\bibnamefont
  {Dal~Corso}}, \bibinfo {author} {\bibfnamefont {S.}~\bibnamefont
  {De~Gironcoli}}, \bibinfo {author} {\bibfnamefont {T.}~\bibnamefont
  {Deutsch}}, \bibinfo {author} {\bibfnamefont {J.~K.}\ \bibnamefont
  {Dewhurst}}, \bibinfo {author} {\bibfnamefont {I.}~\bibnamefont {Di~Marco}},
  \bibinfo {author} {\bibfnamefont {C.}~\bibnamefont {Draxl}}, \bibinfo
  {author} {\bibfnamefont {M.}~\bibnamefont {Du\l{}ak}}, \bibinfo {author}
  {\bibfnamefont {O.}~\bibnamefont {Eriksson}}, \bibinfo {author}
  {\bibfnamefont {J.~A.}\ \bibnamefont {Flores-Livas}}, \bibinfo {author}
  {\bibfnamefont {K.~F.}\ \bibnamefont {Garrity}}, \bibinfo {author}
  {\bibfnamefont {L.}~\bibnamefont {Genovese}}, \bibinfo {author}
  {\bibfnamefont {P.}~\bibnamefont {Giannozzi}}, \bibinfo {author}
  {\bibfnamefont {M.}~\bibnamefont {Giantomassi}}, \bibinfo {author}
  {\bibfnamefont {S.}~\bibnamefont {Goedecker}}, \bibinfo {author}
  {\bibfnamefont {X.}~\bibnamefont {Gonze}}, \bibinfo {author} {\bibfnamefont
  {O.}~\bibnamefont {Gr\r{a}n\"{a}s}}, \bibinfo {author} {\bibfnamefont
  {E.}~\bibnamefont {Gross}}, \bibinfo {author} {\bibfnamefont
  {A.}~\bibnamefont {Gulans}}, \bibinfo {author} {\bibfnamefont
  {F.}~\bibnamefont {Gygi}}, \bibinfo {author} {\bibfnamefont {D.}~\bibnamefont
  {Hamann}}, \bibinfo {author} {\bibfnamefont {P.~J.}\ \bibnamefont {Hasnip}},
  \bibinfo {author} {\bibfnamefont {N.}~\bibnamefont {Holzwarth}}, \bibinfo
  {author} {\bibfnamefont {D.}~\bibnamefont {Iu\c{s}an}}, \bibinfo {author}
  {\bibfnamefont {D.~B.}\ \bibnamefont {Jochym}}, \bibinfo {author}
  {\bibfnamefont {F.}~\bibnamefont {Jollet}}, \bibinfo {author} {\bibfnamefont
  {D.}~\bibnamefont {Jones}}, \bibinfo {author} {\bibfnamefont
  {G.}~\bibnamefont {Kresse}}, \bibinfo {author} {\bibfnamefont
  {K.}~\bibnamefont {Koepernik}}, \bibinfo {author} {\bibfnamefont
  {E.}~\bibnamefont {K\"{u}\c{c}\"{u}kbenli}}, \bibinfo {author} {\bibfnamefont
  {Y.~O.}\ \bibnamefont {Kvashnin}}, \bibinfo {author} {\bibfnamefont {I.~L.}\
  \bibnamefont {Locht}}, \bibinfo {author} {\bibfnamefont {S.}~\bibnamefont
  {Lubeck}}, \bibinfo {author} {\bibfnamefont {M.}~\bibnamefont {Marsman}},
  \bibinfo {author} {\bibfnamefont {N.}~\bibnamefont {Marzari}}, \bibinfo
  {author} {\bibfnamefont {U.}~\bibnamefont {Nitzsche}}, \bibinfo {author}
  {\bibfnamefont {L.}~\bibnamefont {Nordstr\"{o}m}}, \bibinfo {author}
  {\bibfnamefont {T.}~\bibnamefont {Ozaki}}, \bibinfo {author} {\bibfnamefont
  {L.}~\bibnamefont {Paulatto}}, \bibinfo {author} {\bibfnamefont {C.~J.}\
  \bibnamefont {Pickard}}, \bibinfo {author} {\bibfnamefont {W.}~\bibnamefont
  {Poelmans}}, \bibinfo {author} {\bibfnamefont {M.~I.}\ \bibnamefont
  {Probert}}, \bibinfo {author} {\bibfnamefont {K.}~\bibnamefont {Refson}},
  \bibinfo {author} {\bibfnamefont {M.}~\bibnamefont {Richter}}, \bibinfo
  {author} {\bibfnamefont {G.-M.}\ \bibnamefont {Rignanese}}, \bibinfo {author}
  {\bibfnamefont {S.}~\bibnamefont {Saha}}, \bibinfo {author} {\bibfnamefont
  {M.}~\bibnamefont {Scheffler}}, \bibinfo {author} {\bibfnamefont
  {M.}~\bibnamefont {Schlipf}}, \bibinfo {author} {\bibfnamefont
  {K.}~\bibnamefont {Schwarz}}, \bibinfo {author} {\bibfnamefont
  {S.}~\bibnamefont {Sharma}}, \bibinfo {author} {\bibfnamefont
  {F.}~\bibnamefont {Tavazza}}, \bibinfo {author} {\bibfnamefont
  {P.}~\bibnamefont {Thunstr\"{o}m}}, \bibinfo {author} {\bibfnamefont
  {A.}~\bibnamefont {Tkatchenko}}, \bibinfo {author} {\bibfnamefont
  {M.}~\bibnamefont {Torrent}}, \bibinfo {author} {\bibfnamefont
  {D.}~\bibnamefont {Vanderbilt}}, \bibinfo {author} {\bibfnamefont {M.~J.}\
  \bibnamefont {Van~Setten}}, \bibinfo {author} {\bibfnamefont
  {V.}~\bibnamefont {Van~Speybroeck}}, \bibinfo {author} {\bibfnamefont
  {J.~M.}\ \bibnamefont {Wills}}, \bibinfo {author} {\bibfnamefont {J.~R.}\
  \bibnamefont {Yates}}, \bibinfo {author} {\bibfnamefont {G.-X.}\ \bibnamefont
  {Zhang}},\ and\ \bibinfo {author} {\bibfnamefont {S.}~\bibnamefont
  {Cottenier}},\ }\bibfield  {title} {\bibinfo {title} {{Reproducibility in
  density functional theory calculations of solids}},\ }\href
  {https://doi.org/10.1126/science.aad3000} {\bibfield  {journal} {\bibinfo
  {journal} {Science}\ }\textbf {\bibinfo {volume} {351}},\ \bibinfo {pages}
  {aad3000} (\bibinfo {year} {2016})}\BibitemShut {NoStop}%
\bibitem [{\citenamefont {von Barth}\ and\ \citenamefont
  {Hedin}(1972)}]{vonBarth_J_Phys_C_5_1629_1972}%
  \BibitemOpen
  \bibfield  {author} {\bibinfo {author} {\bibfnamefont {U.}~\bibnamefont {von
  Barth}}\ and\ \bibinfo {author} {\bibfnamefont {L.}~\bibnamefont {Hedin}},\
  }\bibfield  {title} {\bibinfo {title} {{A local exchange-correlation
  potential for the spin polarized case. i}},\ }\href
  {https://doi.org/10.1088/0022-3719/5/13/012} {\bibfield  {journal} {\bibinfo
  {journal} {J. Phys. C: Solid State Phys.}\ }\textbf {\bibinfo {volume} {5}},\
  \bibinfo {pages} {1629} (\bibinfo {year} {1972})}\BibitemShut {NoStop}%
\bibitem [{\citenamefont {Kubler}\ \emph {et~al.}(1988)\citenamefont {Kubler},
  \citenamefont {Hock}, \citenamefont {Sticht},\ and\ \citenamefont
  {Williams}}]{Kubler_J_Phys_F_18_469_1988}%
  \BibitemOpen
  \bibfield  {author} {\bibinfo {author} {\bibfnamefont {J.}~\bibnamefont
  {Kubler}}, \bibinfo {author} {\bibfnamefont {K.-H.}\ \bibnamefont {Hock}},
  \bibinfo {author} {\bibfnamefont {J.}~\bibnamefont {Sticht}},\ and\ \bibinfo
  {author} {\bibfnamefont {A.~R.}\ \bibnamefont {Williams}},\ }\bibfield
  {title} {\bibinfo {title} {{Density functional theory of non-collinear
  magnetism}},\ }\href {https://doi.org/10.1088/0305-4608/18/3/018} {\bibfield
  {journal} {\bibinfo  {journal} {J. Phys. F: Met. Phys.}\ }\textbf {\bibinfo
  {volume} {18}},\ \bibinfo {pages} {469} (\bibinfo {year} {1988})}\BibitemShut
  {NoStop}%
\bibitem [{\citenamefont {Perdew}\ \emph {et~al.}(1996)\citenamefont {Perdew},
  \citenamefont {Burke},\ and\ \citenamefont
  {Ernzerhof}}]{Perdew_Phys_Rev_Lett_77_3865_1996}%
  \BibitemOpen
  \bibfield  {author} {\bibinfo {author} {\bibfnamefont {J.~P.}\ \bibnamefont
  {Perdew}}, \bibinfo {author} {\bibfnamefont {K.}~\bibnamefont {Burke}},\ and\
  \bibinfo {author} {\bibfnamefont {M.}~\bibnamefont {Ernzerhof}},\ }\bibfield
  {title} {\bibinfo {title} {{Generalized Gradient Approximation Made
  Simple}},\ }\href {https://doi.org/10.1103/PhysRevLett.77.3865} {\bibfield
  {journal} {\bibinfo  {journal} {Phys. Rev. Lett.}\ }\textbf {\bibinfo
  {volume} {77}},\ \bibinfo {pages} {3865} (\bibinfo {year}
  {1996})}\BibitemShut {NoStop}%
\bibitem [{\citenamefont {Mostofi}\ \emph {et~al.}(2008)\citenamefont
  {Mostofi}, \citenamefont {Yates}, \citenamefont {Lee}, \citenamefont {Souza},
  \citenamefont {Vanderbilt},\ and\ \citenamefont
  {Marzari}}]{Arash_Comput_Phys_Commun_178_685_2008}%
  \BibitemOpen
  \bibfield  {author} {\bibinfo {author} {\bibfnamefont {A.~A.}\ \bibnamefont
  {Mostofi}}, \bibinfo {author} {\bibfnamefont {J.~R.}\ \bibnamefont {Yates}},
  \bibinfo {author} {\bibfnamefont {Y.-S.}\ \bibnamefont {Lee}}, \bibinfo
  {author} {\bibfnamefont {I.}~\bibnamefont {Souza}}, \bibinfo {author}
  {\bibfnamefont {D.}~\bibnamefont {Vanderbilt}},\ and\ \bibinfo {author}
  {\bibfnamefont {N.}~\bibnamefont {Marzari}},\ }\bibfield  {title} {\bibinfo
  {title} {{wannier90: A tool for obtaining maximally-localised Wannier
  functions}},\ }\href
  {https://doi.org/https://doi.org/10.1016/j.cpc.2007.11.016} {\bibfield
  {journal} {\bibinfo  {journal} {Comput. Phys. Commun.}\ }\textbf {\bibinfo
  {volume} {178}},\ \bibinfo {pages} {685 } (\bibinfo {year}
  {2008})}\BibitemShut {NoStop}%
\bibitem [{\citenamefont {Mostofi}\ \emph {et~al.}(2014)\citenamefont
  {Mostofi}, \citenamefont {Yates}, \citenamefont {Pizzi}, \citenamefont {Lee},
  \citenamefont {Souza}, \citenamefont {Vanderbilt},\ and\ \citenamefont
  {Marzari}}]{Arash_Comput_Phys_Commun_185_2309_2014}%
  \BibitemOpen
  \bibfield  {author} {\bibinfo {author} {\bibfnamefont {A.~A.}\ \bibnamefont
  {Mostofi}}, \bibinfo {author} {\bibfnamefont {J.~R.}\ \bibnamefont {Yates}},
  \bibinfo {author} {\bibfnamefont {G.}~\bibnamefont {Pizzi}}, \bibinfo
  {author} {\bibfnamefont {Y.-S.}\ \bibnamefont {Lee}}, \bibinfo {author}
  {\bibfnamefont {I.}~\bibnamefont {Souza}}, \bibinfo {author} {\bibfnamefont
  {D.}~\bibnamefont {Vanderbilt}},\ and\ \bibinfo {author} {\bibfnamefont
  {N.}~\bibnamefont {Marzari}},\ }\bibfield  {title} {\bibinfo {title} {{An
  updated version of wannier90: A tool for obtaining maximally-localised
  Wannier functions}},\ }\href
  {https://doi.org/https://doi.org/10.1016/j.cpc.2014.05.003} {\bibfield
  {journal} {\bibinfo  {journal} {Comput. Phys. Commun.}\ }\textbf {\bibinfo
  {volume} {185}},\ \bibinfo {pages} {2309 } (\bibinfo {year}
  {2014})}\BibitemShut {NoStop}%
\bibitem [{\citenamefont {Weng}\ \emph {et~al.}(2009)\citenamefont {Weng},
  \citenamefont {Ozaki},\ and\ \citenamefont
  {Terakura}}]{Weng_Phys_Rev_B_79_235118_2009}%
  \BibitemOpen
  \bibfield  {author} {\bibinfo {author} {\bibfnamefont {H.}~\bibnamefont
  {Weng}}, \bibinfo {author} {\bibfnamefont {T.}~\bibnamefont {Ozaki}},\ and\
  \bibinfo {author} {\bibfnamefont {K.}~\bibnamefont {Terakura}},\ }\bibfield
  {title} {\bibinfo {title} {{Revisiting magnetic coupling in
  transition-metal-benzene complexes with maximally localized Wannier
  functions}},\ }\href {https://doi.org/10.1103/PhysRevB.79.235118} {\bibfield
  {journal} {\bibinfo  {journal} {Phys. Rev. B}\ }\textbf {\bibinfo {volume}
  {79}},\ \bibinfo {pages} {235118} (\bibinfo {year} {2009})}\BibitemShut
  {NoStop}%
\bibitem [{\citenamefont {Pizzi}\ \emph {et~al.}(2014)\citenamefont {Pizzi},
  \citenamefont {Volja}, \citenamefont {Kozinsky}, \citenamefont {Fornari},\
  and\ \citenamefont {Marzari}}]{Pizzi_Comput_Phys_Commun_185_422_2014}%
  \BibitemOpen
  \bibfield  {author} {\bibinfo {author} {\bibfnamefont {G.}~\bibnamefont
  {Pizzi}}, \bibinfo {author} {\bibfnamefont {D.}~\bibnamefont {Volja}},
  \bibinfo {author} {\bibfnamefont {B.}~\bibnamefont {Kozinsky}}, \bibinfo
  {author} {\bibfnamefont {M.}~\bibnamefont {Fornari}},\ and\ \bibinfo {author}
  {\bibfnamefont {N.}~\bibnamefont {Marzari}},\ }\bibfield  {title} {\bibinfo
  {title} {{BoltzWann: A code for the evaluation of thermoelectric and
  electronic transport properties with a maximally-localized Wannier functions
  basis}},\ }\href {https://doi.org/10.1016/j.cpc.2013.09.015} {\bibfield
  {journal} {\bibinfo  {journal} {Comput. Phys. Commun.}\ }\textbf {\bibinfo
  {volume} {185}},\ \bibinfo {pages} {422} (\bibinfo {year}
  {2014})}\BibitemShut {NoStop}%
\bibitem [{\citenamefont {Onoda}\ \emph {et~al.}(2008)\citenamefont {Onoda},
  \citenamefont {Sugimoto},\ and\ \citenamefont
  {Nagaosa}}]{Onoda_Phys_Rev_B_77_165103_2008}%
  \BibitemOpen
  \bibfield  {author} {\bibinfo {author} {\bibfnamefont {S.}~\bibnamefont
  {Onoda}}, \bibinfo {author} {\bibfnamefont {N.}~\bibnamefont {Sugimoto}},\
  and\ \bibinfo {author} {\bibfnamefont {N.}~\bibnamefont {Nagaosa}},\
  }\bibfield  {title} {\bibinfo {title} {{Quantum transport theory of anomalous
  electric, thermoelectric, and thermal Hall effects in ferromagnets}},\ }\href
  {https://doi.org/10.1103/PhysRevB.77.165103} {\bibfield  {journal} {\bibinfo
  {journal} {Phys. Rev. B}\ }\textbf {\bibinfo {volume} {77}},\ \bibinfo
  {pages} {165103} (\bibinfo {year} {2008})}\BibitemShut {NoStop}%
\bibitem [{\citenamefont {Fukui}\ \emph {et~al.}(2005)\citenamefont {Fukui},
  \citenamefont {Hatsugai},\ and\ \citenamefont
  {Suzuki}}]{Fukui_J_Phys_Soc_Jpn_74_1674_2005}%
  \BibitemOpen
  \bibfield  {author} {\bibinfo {author} {\bibfnamefont {T.}~\bibnamefont
  {Fukui}}, \bibinfo {author} {\bibfnamefont {Y.}~\bibnamefont {Hatsugai}},\
  and\ \bibinfo {author} {\bibfnamefont {H.}~\bibnamefont {Suzuki}},\
  }\bibfield  {title} {\bibinfo {title} {{Chern Numbers in Discretized
  Brillouin Zone: Efficient Method of Computing (Spin) Hall Conductances}},\
  }\href {https://doi.org/10.1143/JPSJ.74.1674} {\bibfield  {journal} {\bibinfo
   {journal} {J. Phys. Soc. Jpn.}\ }\textbf {\bibinfo {volume} {74}},\ \bibinfo
  {pages} {1674} (\bibinfo {year} {2005})}\BibitemShut {NoStop}%
\bibitem [{\citenamefont {Sawahata}\ \emph {et~al.}(2018)\citenamefont
  {Sawahata}, \citenamefont {Yamaguchi}, \citenamefont {Kotaka},\ and\
  \citenamefont {Ishii}}]{Sawahata_Jpn_J_Appl_Phys_57_030309_2018}%
  \BibitemOpen
  \bibfield  {author} {\bibinfo {author} {\bibfnamefont {H.}~\bibnamefont
  {Sawahata}}, \bibinfo {author} {\bibfnamefont {N.}~\bibnamefont {Yamaguchi}},
  \bibinfo {author} {\bibfnamefont {H.}~\bibnamefont {Kotaka}},\ and\ \bibinfo
  {author} {\bibfnamefont {F.}~\bibnamefont {Ishii}},\ }\bibfield  {title}
  {\bibinfo {title} {{First-principles study of electric-field-induced
  topological phase transition in one-bilayer Bi(111)}},\ }\href
  {https://doi.org/10.7567/jjap.57.030309} {\bibfield  {journal} {\bibinfo
  {journal} {Jpn. J. Appl. Phys.}\ }\textbf {\bibinfo {volume} {57}},\ \bibinfo
  {pages} {030309} (\bibinfo {year} {2018})}\BibitemShut {NoStop}%
\bibitem [{\citenamefont {Sawahata}\ \emph {et~al.}(2023)\citenamefont
  {Sawahata}, \citenamefont {Yamaguchi}, \citenamefont {Minami},\ and\
  \citenamefont {Ishii}}]{Sawahata_Phys_Rev_B_107_024404_2023}%
  \BibitemOpen
  \bibfield  {author} {\bibinfo {author} {\bibfnamefont {H.}~\bibnamefont
  {Sawahata}}, \bibinfo {author} {\bibfnamefont {N.}~\bibnamefont {Yamaguchi}},
  \bibinfo {author} {\bibfnamefont {S.}~\bibnamefont {Minami}},\ and\ \bibinfo
  {author} {\bibfnamefont {F.}~\bibnamefont {Ishii}},\ }\bibfield  {title}
  {\bibinfo {title} {{First-principles calculation of anomalous Hall and Nernst
  conductivity by local Berry phase}},\ }\href
  {https://doi.org/10.1103/PhysRevB.107.024404} {\bibfield  {journal} {\bibinfo
   {journal} {Phys. Rev. B}\ }\textbf {\bibinfo {volume} {107}},\ \bibinfo
  {pages} {024404} (\bibinfo {year} {2023})}\BibitemShut {NoStop}%
\bibitem [{\citenamefont {Kurz}\ \emph {et~al.}(2004)\citenamefont {Kurz},
  \citenamefont {F\"orster}, \citenamefont {Nordstr\"om}, \citenamefont
  {Bihlmayer},\ and\ \citenamefont
  {Bl\"ugel}}]{Kurz_Phys_Rev_B_69_024415_2004}%
  \BibitemOpen
  \bibfield  {author} {\bibinfo {author} {\bibfnamefont {P.}~\bibnamefont
  {Kurz}}, \bibinfo {author} {\bibfnamefont {F.}~\bibnamefont {F\"orster}},
  \bibinfo {author} {\bibfnamefont {L.}~\bibnamefont {Nordstr\"om}}, \bibinfo
  {author} {\bibfnamefont {G.}~\bibnamefont {Bihlmayer}},\ and\ \bibinfo
  {author} {\bibfnamefont {S.}~\bibnamefont {Bl\"ugel}},\ }\bibfield  {title}
  {\bibinfo {title} {{Ab initio treatment of noncollinear magnets with the
  full-potential linearized augmented plane wave method}},\ }\href
  {https://doi.org/10.1103/PhysRevB.69.024415} {\bibfield  {journal} {\bibinfo
  {journal} {Phys. Rev. B}\ }\textbf {\bibinfo {volume} {69}},\ \bibinfo
  {pages} {024415} (\bibinfo {year} {2004})}\BibitemShut {NoStop}%
\bibitem [{\citenamefont {Ohgushi}\ \emph {et~al.}(2000)\citenamefont
  {Ohgushi}, \citenamefont {Murakami},\ and\ \citenamefont
  {Nagaosa}}]{Ohgushi_Phys_Rev_B_62_R6065_2000}%
  \BibitemOpen
  \bibfield  {author} {\bibinfo {author} {\bibfnamefont {K.}~\bibnamefont
  {Ohgushi}}, \bibinfo {author} {\bibfnamefont {S.}~\bibnamefont {Murakami}},\
  and\ \bibinfo {author} {\bibfnamefont {N.}~\bibnamefont {Nagaosa}},\
  }\bibfield  {title} {\bibinfo {title} {{Spin anisotropy and quantum Hall
  effect in the kagom\'e lattice: Chiral spin state based on a ferromagnet}},\
  }\href {https://doi.org/10.1103/PhysRevB.62.R6065} {\bibfield  {journal}
  {\bibinfo  {journal} {Phys. Rev. B}\ }\textbf {\bibinfo {volume} {62}},\
  \bibinfo {pages} {R6065} (\bibinfo {year} {2000})}\BibitemShut {NoStop}%
\bibitem [{\citenamefont {Munakata}\ \emph {et~al.}(1992)\citenamefont
  {Munakata}, \citenamefont {Matsuura}, \citenamefont {Kubo}, \citenamefont
  {Kawano},\ and\ \citenamefont
  {Yamauchi}}]{Munakata_Phys_Rev_B_45_10604_1992}%
  \BibitemOpen
  \bibfield  {author} {\bibinfo {author} {\bibfnamefont {F.}~\bibnamefont
  {Munakata}}, \bibinfo {author} {\bibfnamefont {K.}~\bibnamefont {Matsuura}},
  \bibinfo {author} {\bibfnamefont {K.}~\bibnamefont {Kubo}}, \bibinfo {author}
  {\bibfnamefont {T.}~\bibnamefont {Kawano}},\ and\ \bibinfo {author}
  {\bibfnamefont {H.}~\bibnamefont {Yamauchi}},\ }\bibfield  {title} {\bibinfo
  {title} {{Thermoelectric power of
  ${\mathrm{Bi}}_{2}$${\mathrm{Sr}}_{2}$${\mathrm{Ca}}_{1\mathrm{\ensuremath{-}}\mathit{x}}$${\mathrm{Y}}_{\mathit{x}}$${\mathrm{Cu}}_{2}$${\mathrm{O}}_{8+\mathit{y}}$}},\
  }\href {https://doi.org/10.1103/PhysRevB.45.10604} {\bibfield  {journal}
  {\bibinfo  {journal} {Phys. Rev. B}\ }\textbf {\bibinfo {volume} {45}},\
  \bibinfo {pages} {10604} (\bibinfo {year} {1992})}\BibitemShut {NoStop}%
\bibitem [{\citenamefont {Yanagi}\ \emph {et~al.}(2014)\citenamefont {Yanagi},
  \citenamefont {Kanda}, \citenamefont {Oshima}, \citenamefont {Kitamura},
  \citenamefont {Kawai}, \citenamefont {Yamamoto}, \citenamefont {Takenobu},
  \citenamefont {Nakai},\ and\ \citenamefont
  {Maniwa}}]{Yanagi_Nano_Lett_14_6437_2014}%
  \BibitemOpen
  \bibfield  {author} {\bibinfo {author} {\bibfnamefont {K.}~\bibnamefont
  {Yanagi}}, \bibinfo {author} {\bibfnamefont {S.}~\bibnamefont {Kanda}},
  \bibinfo {author} {\bibfnamefont {Y.}~\bibnamefont {Oshima}}, \bibinfo
  {author} {\bibfnamefont {Y.}~\bibnamefont {Kitamura}}, \bibinfo {author}
  {\bibfnamefont {H.}~\bibnamefont {Kawai}}, \bibinfo {author} {\bibfnamefont
  {T.}~\bibnamefont {Yamamoto}}, \bibinfo {author} {\bibfnamefont
  {T.}~\bibnamefont {Takenobu}}, \bibinfo {author} {\bibfnamefont
  {Y.}~\bibnamefont {Nakai}},\ and\ \bibinfo {author} {\bibfnamefont
  {Y.}~\bibnamefont {Maniwa}},\ }\bibfield  {title} {\bibinfo {title} {{Tuning
  of the Thermoelectric Properties of One-Dimensional Material Networks by
  Electric Double Layer Techniques Using Ionic Liquids}},\ }\href
  {https://doi.org/10.1021/nl502982f} {\bibfield  {journal} {\bibinfo
  {journal} {Nano Lett.}\ }\textbf {\bibinfo {volume} {14}},\ \bibinfo {pages}
  {6437} (\bibinfo {year} {2014})}\BibitemShut {NoStop}%
\bibitem [{\citenamefont {Newns}\ \emph {et~al.}(1994)\citenamefont {Newns},
  \citenamefont {Tsuei}, \citenamefont {Huebener}, \citenamefont {van Bentum},
  \citenamefont {Pattnaik},\ and\ \citenamefont
  {Chi}}]{Newns_Phys_Rev_Lett_73_1695_1994}%
  \BibitemOpen
  \bibfield  {author} {\bibinfo {author} {\bibfnamefont {D.~M.}\ \bibnamefont
  {Newns}}, \bibinfo {author} {\bibfnamefont {C.~C.}\ \bibnamefont {Tsuei}},
  \bibinfo {author} {\bibfnamefont {R.~P.}\ \bibnamefont {Huebener}}, \bibinfo
  {author} {\bibfnamefont {P.~J.~M.}\ \bibnamefont {van Bentum}}, \bibinfo
  {author} {\bibfnamefont {P.~C.}\ \bibnamefont {Pattnaik}},\ and\ \bibinfo
  {author} {\bibfnamefont {C.~C.}\ \bibnamefont {Chi}},\ }\bibfield  {title}
  {\bibinfo {title} {{Quasiclassical Transport at a van Hove Singularity in
  Cuprate Superconductors}},\ }\href
  {https://doi.org/10.1103/PhysRevLett.73.1695} {\bibfield  {journal} {\bibinfo
   {journal} {Phys. Rev. Lett.}\ }\textbf {\bibinfo {volume} {73}},\ \bibinfo
  {pages} {1695} (\bibinfo {year} {1994})}\BibitemShut {NoStop}%
\bibitem [{\citenamefont {Zhu}\ \emph {et~al.}(2020)\citenamefont {Zhu},
  \citenamefont {Yao}, \citenamefont {Jiang},\ and\ \citenamefont
  {Zheng}}]{Zhu_Appl_Phys_Lett_116_022404_2020}%
  \BibitemOpen
  \bibfield  {author} {\bibinfo {author} {\bibfnamefont {M.}~\bibnamefont
  {Zhu}}, \bibinfo {author} {\bibfnamefont {H.}~\bibnamefont {Yao}}, \bibinfo
  {author} {\bibfnamefont {L.}~\bibnamefont {Jiang}},\ and\ \bibinfo {author}
  {\bibfnamefont {Y.}~\bibnamefont {Zheng}},\ }\bibfield  {title} {\bibinfo
  {title} {{Theoretical model of spintronic device based on tunable anomalous
  Hall conductivity of monolayer CrI$_3$}},\ }\href
  {https://doi.org/10.1063/1.5132356} {\bibfield  {journal} {\bibinfo
  {journal} {Appl. Phys. Lett.}\ }\textbf {\bibinfo {volume} {116}},\ \bibinfo
  {pages} {022404} (\bibinfo {year} {2020})}\BibitemShut {NoStop}%
\bibitem [{\citenamefont {Sakai}\ \emph {et~al.}(2020)\citenamefont {Sakai},
  \citenamefont {Minami}, \citenamefont {Koretsune}, \citenamefont {Chen},
  \citenamefont {Higo}, \citenamefont {Wang}, \citenamefont {Nomoto},
  \citenamefont {Hirayama}, \citenamefont {Miwa}, \citenamefont
  {Nishio-Hamane}, \citenamefont {Ishii}, \citenamefont {Arita},\ and\
  \citenamefont {Nakatsuji}}]{Sakai_Nature_581_53_2020}%
  \BibitemOpen
  \bibfield  {author} {\bibinfo {author} {\bibfnamefont {A.}~\bibnamefont
  {Sakai}}, \bibinfo {author} {\bibfnamefont {S.}~\bibnamefont {Minami}},
  \bibinfo {author} {\bibfnamefont {T.}~\bibnamefont {Koretsune}}, \bibinfo
  {author} {\bibfnamefont {T.}~\bibnamefont {Chen}}, \bibinfo {author}
  {\bibfnamefont {T.}~\bibnamefont {Higo}}, \bibinfo {author} {\bibfnamefont
  {Y.}~\bibnamefont {Wang}}, \bibinfo {author} {\bibfnamefont {T.}~\bibnamefont
  {Nomoto}}, \bibinfo {author} {\bibfnamefont {M.}~\bibnamefont {Hirayama}},
  \bibinfo {author} {\bibfnamefont {S.}~\bibnamefont {Miwa}}, \bibinfo {author}
  {\bibfnamefont {D.}~\bibnamefont {Nishio-Hamane}}, \bibinfo {author}
  {\bibfnamefont {F.}~\bibnamefont {Ishii}}, \bibinfo {author} {\bibfnamefont
  {R.}~\bibnamefont {Arita}},\ and\ \bibinfo {author} {\bibfnamefont
  {S.}~\bibnamefont {Nakatsuji}},\ }\bibfield  {title} {\bibinfo {title}
  {{Iron-based binary ferromagnets for transverse thermoelectric conversion}},\
  }\href {https://www.nature.com/articles/s41586-020-2230-z} {\bibfield
  {journal} {\bibinfo  {journal} {Nature}\ }\textbf {\bibinfo {volume} {581}},\
  \bibinfo {pages} {53} (\bibinfo {year} {2020})}\BibitemShut {NoStop}%
\bibitem [{\citenamefont {Bonilla}\ \emph {et~al.}(2018)\citenamefont
  {Bonilla}, \citenamefont {Kolekar}, \citenamefont {Ma}, \citenamefont {Diaz},
  \citenamefont {Kalappattil}, \citenamefont {Das}, \citenamefont {Eggers},
  \citenamefont {Gutierrez}, \citenamefont {Phan},\ and\ \citenamefont
  {Batzill}}]{Bonilla_Nat_Nanotechnol_13_289_2018}%
  \BibitemOpen
  \bibfield  {author} {\bibinfo {author} {\bibfnamefont {M.}~\bibnamefont
  {Bonilla}}, \bibinfo {author} {\bibfnamefont {S.}~\bibnamefont {Kolekar}},
  \bibinfo {author} {\bibfnamefont {Y.}~\bibnamefont {Ma}}, \bibinfo {author}
  {\bibfnamefont {H.~C.}\ \bibnamefont {Diaz}}, \bibinfo {author}
  {\bibfnamefont {V.}~\bibnamefont {Kalappattil}}, \bibinfo {author}
  {\bibfnamefont {R.}~\bibnamefont {Das}}, \bibinfo {author} {\bibfnamefont
  {T.}~\bibnamefont {Eggers}}, \bibinfo {author} {\bibfnamefont {H.~R.}\
  \bibnamefont {Gutierrez}}, \bibinfo {author} {\bibfnamefont {M.-H.}\
  \bibnamefont {Phan}},\ and\ \bibinfo {author} {\bibfnamefont
  {M.}~\bibnamefont {Batzill}},\ }\bibfield  {title} {\bibinfo {title} {{Strong
  room-temperature ferromagnetism in VSe$_2$ monolayers on van der Waals
  substrates}},\ }\href {https://doi.org/10.1038/s41565-018-0063-9} {\bibfield
  {journal} {\bibinfo  {journal} {Nat. Nanotechnol.}\ }\textbf {\bibinfo
  {volume} {13}},\ \bibinfo {pages} {289} (\bibinfo {year} {2018})}\BibitemShut
  {NoStop}%
\bibitem [{\citenamefont {Yu}\ \emph {et~al.}(2019)\citenamefont {Yu},
  \citenamefont {Li}, \citenamefont {Herng}, \citenamefont {Wang},
  \citenamefont {Zhao}, \citenamefont {Chi}, \citenamefont {Fu}, \citenamefont
  {Abdelwahab}, \citenamefont {Zhou}, \citenamefont {Dan}, \citenamefont
  {Chen}, \citenamefont {Chen}, \citenamefont {Li}, \citenamefont {Lu},
  \citenamefont {Pennycook}, \citenamefont {Feng}, \citenamefont {Ding},\ and\
  \citenamefont {Loh}}]{Yu_Adv_Mater_31_1903779_2019}%
  \BibitemOpen
  \bibfield  {author} {\bibinfo {author} {\bibfnamefont {W.}~\bibnamefont
  {Yu}}, \bibinfo {author} {\bibfnamefont {J.}~\bibnamefont {Li}}, \bibinfo
  {author} {\bibfnamefont {T.~S.}\ \bibnamefont {Herng}}, \bibinfo {author}
  {\bibfnamefont {Z.}~\bibnamefont {Wang}}, \bibinfo {author} {\bibfnamefont
  {X.}~\bibnamefont {Zhao}}, \bibinfo {author} {\bibfnamefont {X.}~\bibnamefont
  {Chi}}, \bibinfo {author} {\bibfnamefont {W.}~\bibnamefont {Fu}}, \bibinfo
  {author} {\bibfnamefont {I.}~\bibnamefont {Abdelwahab}}, \bibinfo {author}
  {\bibfnamefont {J.}~\bibnamefont {Zhou}}, \bibinfo {author} {\bibfnamefont
  {J.}~\bibnamefont {Dan}}, \bibinfo {author} {\bibfnamefont {Z.}~\bibnamefont
  {Chen}}, \bibinfo {author} {\bibfnamefont {Z.}~\bibnamefont {Chen}}, \bibinfo
  {author} {\bibfnamefont {Z.}~\bibnamefont {Li}}, \bibinfo {author}
  {\bibfnamefont {J.}~\bibnamefont {Lu}}, \bibinfo {author} {\bibfnamefont
  {S.~J.}\ \bibnamefont {Pennycook}}, \bibinfo {author} {\bibfnamefont {Y.~P.}\
  \bibnamefont {Feng}}, \bibinfo {author} {\bibfnamefont {J.}~\bibnamefont
  {Ding}},\ and\ \bibinfo {author} {\bibfnamefont {K.~P.}\ \bibnamefont
  {Loh}},\ }\bibfield  {title} {\bibinfo {title} {{Chemically Exfoliated
  VSe$_2$ Monolayers with Room-Temperature Ferromagnetism}},\ }\href
  {https://doi.org/10.1002/adma.201903779} {\bibfield  {journal} {\bibinfo
  {journal} {Adv. Mater.}\ }\textbf {\bibinfo {volume} {31}},\ \bibinfo {pages}
  {1903779} (\bibinfo {year} {2019})}\BibitemShut {NoStop}%
\bibitem [{\citenamefont {Chua}\ \emph {et~al.}(2021)\citenamefont {Chua},
  \citenamefont {Zhou}, \citenamefont {Yu}, \citenamefont {Yu}, \citenamefont
  {Gou}, \citenamefont {Zhu}, \citenamefont {Zhang}, \citenamefont {Liu},
  \citenamefont {Breese}, \citenamefont {Chen}, \citenamefont {Loh},
  \citenamefont {Feng}, \citenamefont {Yang}, \citenamefont {Huang},\ and\
  \citenamefont {Wee}}]{Chua_Adv_Mater_33_2103360_2021}%
  \BibitemOpen
  \bibfield  {author} {\bibinfo {author} {\bibfnamefont {R.}~\bibnamefont
  {Chua}}, \bibinfo {author} {\bibfnamefont {J.}~\bibnamefont {Zhou}}, \bibinfo
  {author} {\bibfnamefont {X.}~\bibnamefont {Yu}}, \bibinfo {author}
  {\bibfnamefont {W.}~\bibnamefont {Yu}}, \bibinfo {author} {\bibfnamefont
  {J.}~\bibnamefont {Gou}}, \bibinfo {author} {\bibfnamefont {R.}~\bibnamefont
  {Zhu}}, \bibinfo {author} {\bibfnamefont {L.}~\bibnamefont {Zhang}}, \bibinfo
  {author} {\bibfnamefont {M.}~\bibnamefont {Liu}}, \bibinfo {author}
  {\bibfnamefont {M.~B.~H.}\ \bibnamefont {Breese}}, \bibinfo {author}
  {\bibfnamefont {W.}~\bibnamefont {Chen}}, \bibinfo {author} {\bibfnamefont
  {K.~P.}\ \bibnamefont {Loh}}, \bibinfo {author} {\bibfnamefont {Y.~P.}\
  \bibnamefont {Feng}}, \bibinfo {author} {\bibfnamefont {M.}~\bibnamefont
  {Yang}}, \bibinfo {author} {\bibfnamefont {Y.~L.}\ \bibnamefont {Huang}},\
  and\ \bibinfo {author} {\bibfnamefont {A.~T.~S.}\ \bibnamefont {Wee}},\
  }\bibfield  {title} {\bibinfo {title} {{Room Temperature Ferromagnetism of
  Monolayer Chromium Telluride with Perpendicular Magnetic Anisotropy}},\
  }\href {https://doi.org/10.1002/adma.202103360} {\bibfield  {journal}
  {\bibinfo  {journal} {Adv. Mater.}\ }\textbf {\bibinfo {volume} {33}},\
  \bibinfo {pages} {2103360} (\bibinfo {year} {2021})}\BibitemShut {NoStop}%
\bibitem [{\citenamefont {Wu}\ \emph {et~al.}(2021)\citenamefont {Wu},
  \citenamefont {Zhang}, \citenamefont {Yang}, \citenamefont {Wang},
  \citenamefont {Li}, \citenamefont {Li}, \citenamefont {Gao}, \citenamefont
  {Zhang}, \citenamefont {Du}, \citenamefont {Shu},\ and\ \citenamefont
  {Chang}}]{Wu_Nat_Commun_12_5688_2021}%
  \BibitemOpen
  \bibfield  {author} {\bibinfo {author} {\bibfnamefont {H.}~\bibnamefont
  {Wu}}, \bibinfo {author} {\bibfnamefont {W.}~\bibnamefont {Zhang}}, \bibinfo
  {author} {\bibfnamefont {L.}~\bibnamefont {Yang}}, \bibinfo {author}
  {\bibfnamefont {J.}~\bibnamefont {Wang}}, \bibinfo {author} {\bibfnamefont
  {J.}~\bibnamefont {Li}}, \bibinfo {author} {\bibfnamefont {L.}~\bibnamefont
  {Li}}, \bibinfo {author} {\bibfnamefont {Y.}~\bibnamefont {Gao}}, \bibinfo
  {author} {\bibfnamefont {L.}~\bibnamefont {Zhang}}, \bibinfo {author}
  {\bibfnamefont {J.}~\bibnamefont {Du}}, \bibinfo {author} {\bibfnamefont
  {H.}~\bibnamefont {Shu}},\ and\ \bibinfo {author} {\bibfnamefont
  {H.}~\bibnamefont {Chang}},\ }\bibfield  {title} {\bibinfo {title} {{Strong
  intrinsic room-temperature ferromagnetism in freestanding non-van der Waals
  ultrathin 2D crystals}},\ }\href {https://doi.org/10.1038/s41467-021-26009-0}
  {\bibfield  {journal} {\bibinfo  {journal} {Nat. Commun.}\ }\textbf {\bibinfo
  {volume} {12}},\ \bibinfo {pages} {5688} (\bibinfo {year}
  {2021})}\BibitemShut {NoStop}%
\bibitem [{\citenamefont {O'Hara}\ \emph {et~al.}(2018)\citenamefont {O'Hara},
  \citenamefont {Zhu}, \citenamefont {Trout}, \citenamefont {Ahmed},
  \citenamefont {Luo}, \citenamefont {Lee}, \citenamefont {Brenner},
  \citenamefont {Rajan}, \citenamefont {Gupta}, \citenamefont {McComb},\ and\
  \citenamefont {Kawakami}}]{OHara_Nano_Lett_18_3125_2018}%
  \BibitemOpen
  \bibfield  {author} {\bibinfo {author} {\bibfnamefont {D.~J.}\ \bibnamefont
  {O'Hara}}, \bibinfo {author} {\bibfnamefont {T.}~\bibnamefont {Zhu}},
  \bibinfo {author} {\bibfnamefont {A.~H.}\ \bibnamefont {Trout}}, \bibinfo
  {author} {\bibfnamefont {A.~S.}\ \bibnamefont {Ahmed}}, \bibinfo {author}
  {\bibfnamefont {Y.~K.}\ \bibnamefont {Luo}}, \bibinfo {author} {\bibfnamefont
  {C.~H.}\ \bibnamefont {Lee}}, \bibinfo {author} {\bibfnamefont {M.~R.}\
  \bibnamefont {Brenner}}, \bibinfo {author} {\bibfnamefont {S.}~\bibnamefont
  {Rajan}}, \bibinfo {author} {\bibfnamefont {J.~A.}\ \bibnamefont {Gupta}},
  \bibinfo {author} {\bibfnamefont {D.~W.}\ \bibnamefont {McComb}},\ and\
  \bibinfo {author} {\bibfnamefont {R.~K.}\ \bibnamefont {Kawakami}},\
  }\bibfield  {title} {\bibinfo {title} {{Room Temperature Intrinsic
  Ferromagnetism in Epitaxial Manganese Selenide Films in the Monolayer
  Limit}},\ }\href {https://doi.org/10.1021/acs.nanolett.8b00683} {\bibfield
  {journal} {\bibinfo  {journal} {Nano Lett.}\ }\textbf {\bibinfo {volume}
  {18}},\ \bibinfo {pages} {3125} (\bibinfo {year} {2018})}\BibitemShut
  {NoStop}%
\bibitem [{\citenamefont {Zhang}\ \emph {et~al.}(2022)\citenamefont {Zhang},
  \citenamefont {Guo}, \citenamefont {Wu}, \citenamefont {Wen}, \citenamefont
  {Yang}, \citenamefont {Jin}, \citenamefont {Zhang},\ and\ \citenamefont
  {Chang}}]{Zhang_Nat_Commun_13_5067_2022}%
  \BibitemOpen
  \bibfield  {author} {\bibinfo {author} {\bibfnamefont {G.}~\bibnamefont
  {Zhang}}, \bibinfo {author} {\bibfnamefont {F.}~\bibnamefont {Guo}}, \bibinfo
  {author} {\bibfnamefont {H.}~\bibnamefont {Wu}}, \bibinfo {author}
  {\bibfnamefont {X.}~\bibnamefont {Wen}}, \bibinfo {author} {\bibfnamefont
  {L.}~\bibnamefont {Yang}}, \bibinfo {author} {\bibfnamefont {W.}~\bibnamefont
  {Jin}}, \bibinfo {author} {\bibfnamefont {W.}~\bibnamefont {Zhang}},\ and\
  \bibinfo {author} {\bibfnamefont {H.}~\bibnamefont {Chang}},\ }\bibfield
  {title} {\bibinfo {title} {{Above-room-temperature strong intrinsic
  ferromagnetism in 2D van der Waals Fe$_3$GaTe$_2$ with large perpendicular
  magnetic anisotropy}},\ }\href {https://doi.org/10.1038/s41467-022-32605-5}
  {\bibfield  {journal} {\bibinfo  {journal} {Nat. Commun.}\ }\textbf {\bibinfo
  {volume} {13}},\ \bibinfo {pages} {5067} (\bibinfo {year}
  {2022})}\BibitemShut {NoStop}%
\bibitem [{\citenamefont {Yuan}\ \emph {et~al.}(2019)\citenamefont {Yuan},
  \citenamefont {Balk}, \citenamefont {Guo}, \citenamefont {Fang},
  \citenamefont {Patel}, \citenamefont {Zhao}, \citenamefont {Terlier},
  \citenamefont {Natelson}, \citenamefont {Crooker},\ and\ \citenamefont
  {Lou}}]{Yuan_Nano_Lett_19_3777_2019}%
  \BibitemOpen
  \bibfield  {author} {\bibinfo {author} {\bibfnamefont {J.}~\bibnamefont
  {Yuan}}, \bibinfo {author} {\bibfnamefont {A.}~\bibnamefont {Balk}}, \bibinfo
  {author} {\bibfnamefont {H.}~\bibnamefont {Guo}}, \bibinfo {author}
  {\bibfnamefont {Q.}~\bibnamefont {Fang}}, \bibinfo {author} {\bibfnamefont
  {S.}~\bibnamefont {Patel}}, \bibinfo {author} {\bibfnamefont
  {X.}~\bibnamefont {Zhao}}, \bibinfo {author} {\bibfnamefont {T.}~\bibnamefont
  {Terlier}}, \bibinfo {author} {\bibfnamefont {D.}~\bibnamefont {Natelson}},
  \bibinfo {author} {\bibfnamefont {S.}~\bibnamefont {Crooker}},\ and\ \bibinfo
  {author} {\bibfnamefont {J.}~\bibnamefont {Lou}},\ }\bibfield  {title}
  {\bibinfo {title} {{Room-Temperature Magnetic Order in Air-Stable Ultrathin
  Iron Oxide}},\ }\href {https://doi.org/10.1021/acs.nanolett.9b00905}
  {\bibfield  {journal} {\bibinfo  {journal} {Nano Lett.}\ }\textbf {\bibinfo
  {volume} {19}},\ \bibinfo {pages} {3777} (\bibinfo {year}
  {2019})}\BibitemShut {NoStop}%
\bibitem [{\citenamefont {Cheng}\ \emph {et~al.}(2022)\citenamefont {Cheng},
  \citenamefont {Yin}, \citenamefont {Wen}, \citenamefont {Zhai}, \citenamefont
  {Guo}, \citenamefont {Zhang}, \citenamefont {Liao}, \citenamefont {Xiong},
  \citenamefont {Wang}, \citenamefont {Yuan}, \citenamefont {Jiang},
  \citenamefont {Liu},\ and\ \citenamefont
  {He}}]{Cheng_Nat_Commun_13_5241_2022}%
  \BibitemOpen
  \bibfield  {author} {\bibinfo {author} {\bibfnamefont {R.}~\bibnamefont
  {Cheng}}, \bibinfo {author} {\bibfnamefont {L.}~\bibnamefont {Yin}}, \bibinfo
  {author} {\bibfnamefont {Y.}~\bibnamefont {Wen}}, \bibinfo {author}
  {\bibfnamefont {B.}~\bibnamefont {Zhai}}, \bibinfo {author} {\bibfnamefont
  {Y.}~\bibnamefont {Guo}}, \bibinfo {author} {\bibfnamefont {Z.}~\bibnamefont
  {Zhang}}, \bibinfo {author} {\bibfnamefont {W.}~\bibnamefont {Liao}},
  \bibinfo {author} {\bibfnamefont {W.}~\bibnamefont {Xiong}}, \bibinfo
  {author} {\bibfnamefont {H.}~\bibnamefont {Wang}}, \bibinfo {author}
  {\bibfnamefont {S.}~\bibnamefont {Yuan}}, \bibinfo {author} {\bibfnamefont
  {J.}~\bibnamefont {Jiang}}, \bibinfo {author} {\bibfnamefont
  {C.}~\bibnamefont {Liu}},\ and\ \bibinfo {author} {\bibfnamefont
  {J.}~\bibnamefont {He}},\ }\bibfield  {title} {\bibinfo {title} {{Ultrathin
  ferrite nanosheets for room-temperature two-dimensional magnetic
  semiconductors}},\ }\href {https://doi.org/10.1038/s41467-022-33017-1}
  {\bibfield  {journal} {\bibinfo  {journal} {Nat. Commun.}\ }\textbf {\bibinfo
  {volume} {13}},\ \bibinfo {pages} {5241} (\bibinfo {year}
  {2022})}\BibitemShut {NoStop}%
\bibitem [{\citenamefont {Smrcka}\ and\ \citenamefont
  {Streda}(1977)}]{Smrcka_J_Phys_C_1977}%
  \BibitemOpen
  \bibfield  {author} {\bibinfo {author} {\bibfnamefont {L.}~\bibnamefont
  {Smrcka}}\ and\ \bibinfo {author} {\bibfnamefont {P.}~\bibnamefont
  {Streda}},\ }\bibfield  {title} {\bibinfo {title} {{Transport coefficients in
  strong magnetic fields}},\ }\href
  {https://doi.org/10.1088/0022-3719/10/12/021} {\bibfield  {journal} {\bibinfo
   {journal} {J. Phys. C}\ }\textbf {\bibinfo {volume} {10}},\ \bibinfo {pages}
  {2153} (\bibinfo {year} {1977})}\BibitemShut {NoStop}%
\bibitem [{\citenamefont {Sakai}\ \emph {et~al.}(2018)\citenamefont {Sakai},
  \citenamefont {Mizuta}, \citenamefont {Nugroho}, \citenamefont {Sihombing},
  \citenamefont {Koretsune}, \citenamefont {Suzuki}, \citenamefont {Takemori},
  \citenamefont {Ishii}, \citenamefont {Nishio-Hamane}, \citenamefont {Arita},
  \citenamefont {Goswami},\ and\ \citenamefont
  {Nakatsuji}}]{Sakai_Nat_Phys_14_1119_2018}%
  \BibitemOpen
  \bibfield  {author} {\bibinfo {author} {\bibfnamefont {A.}~\bibnamefont
  {Sakai}}, \bibinfo {author} {\bibfnamefont {Y.~P.}\ \bibnamefont {Mizuta}},
  \bibinfo {author} {\bibfnamefont {A.~A.}\ \bibnamefont {Nugroho}}, \bibinfo
  {author} {\bibfnamefont {R.}~\bibnamefont {Sihombing}}, \bibinfo {author}
  {\bibfnamefont {T.}~\bibnamefont {Koretsune}}, \bibinfo {author}
  {\bibfnamefont {M.}~\bibnamefont {Suzuki}}, \bibinfo {author} {\bibfnamefont
  {N.}~\bibnamefont {Takemori}}, \bibinfo {author} {\bibfnamefont
  {R.}~\bibnamefont {Ishii}}, \bibinfo {author} {\bibfnamefont
  {D.}~\bibnamefont {Nishio-Hamane}}, \bibinfo {author} {\bibfnamefont
  {R.}~\bibnamefont {Arita}}, \bibinfo {author} {\bibfnamefont
  {P.}~\bibnamefont {Goswami}},\ and\ \bibinfo {author} {\bibfnamefont
  {S.}~\bibnamefont {Nakatsuji}},\ }\bibfield  {title} {\bibinfo {title}
  {{Giant anomalous Nernst effect and quantum-critical scaling in a
  ferromagnetic semimetal}},\ }\href
  {https://www.nature.com/articles/s41567-018-0225-6} {\bibfield  {journal}
  {\bibinfo  {journal} {Nat. Phys.}\ }\textbf {\bibinfo {volume} {14}},\
  \bibinfo {pages} {1119} (\bibinfo {year} {2018})}\BibitemShut {NoStop}%
\bibitem [{\citenamefont {Nakamura}\ \emph {et~al.}(2021)\citenamefont
  {Nakamura}, \citenamefont {Minami}, \citenamefont {Tomita}, \citenamefont
  {Nugroho},\ and\ \citenamefont
  {Nakatsuji}}]{Nakamura_Phys_Rev_B_104_L161114_2021}%
  \BibitemOpen
  \bibfield  {author} {\bibinfo {author} {\bibfnamefont {H.}~\bibnamefont
  {Nakamura}}, \bibinfo {author} {\bibfnamefont {S.}~\bibnamefont {Minami}},
  \bibinfo {author} {\bibfnamefont {T.}~\bibnamefont {Tomita}}, \bibinfo
  {author} {\bibfnamefont {A.~A.}\ \bibnamefont {Nugroho}},\ and\ \bibinfo
  {author} {\bibfnamefont {S.}~\bibnamefont {Nakatsuji}},\ }\bibfield  {title}
  {\bibinfo {title} {{Logarithmic criticality in transverse thermoelectric
  conductivity of the ferromagnetic topological semimetal CoMnSb}},\ }\href
  {https://doi.org/10.1103/PhysRevB.104.L161114} {\bibfield  {journal}
  {\bibinfo  {journal} {Phys. Rev. B}\ }\textbf {\bibinfo {volume} {104}},\
  \bibinfo {pages} {L161114} (\bibinfo {year} {2021})}\BibitemShut {NoStop}%
\bibitem [{\citenamefont {Smit}(1955)}]{Smit_Physica_21_877_1955}%
  \BibitemOpen
  \bibfield  {author} {\bibinfo {author} {\bibfnamefont {J.}~\bibnamefont
  {Smit}},\ }\bibfield  {title} {\bibinfo {title} {{The spontaneous Hall effect
  in ferromagnetics I}},\ }\href
  {https://doi.org/10.1016/S0031-8914(55)92596-9} {\bibfield  {journal}
  {\bibinfo  {journal} {Physica}\ }\textbf {\bibinfo {volume} {21}},\ \bibinfo
  {pages} {877} (\bibinfo {year} {1955})}\BibitemShut {NoStop}%
\bibitem [{\citenamefont {Smit}(1958)}]{Smit_Physica_24_39_1958}%
  \BibitemOpen
  \bibfield  {author} {\bibinfo {author} {\bibfnamefont {J.}~\bibnamefont
  {Smit}},\ }\bibfield  {title} {\bibinfo {title} {{The spontaneous Hall effect
  in ferromagnetics II}},\ }\href
  {https://doi.org/10.1016/S0031-8914(58)93541-9} {\bibfield  {journal}
  {\bibinfo  {journal} {Physica}\ }\textbf {\bibinfo {volume} {24}},\ \bibinfo
  {pages} {39} (\bibinfo {year} {1958})}\BibitemShut {NoStop}%
\bibitem [{\citenamefont {Berger}(1970)}]{Berger_Phys_Rev_B_2_4559_1970}%
  \BibitemOpen
  \bibfield  {author} {\bibinfo {author} {\bibfnamefont {L.}~\bibnamefont
  {Berger}},\ }\bibfield  {title} {\bibinfo {title} {{Side-Jump Mechanism for
  the Hall Effect of Ferromagnets}},\ }\href
  {https://doi.org/10.1103/PhysRevB.2.4559} {\bibfield  {journal} {\bibinfo
  {journal} {Phys. Rev. B}\ }\textbf {\bibinfo {volume} {2}},\ \bibinfo {pages}
  {4559} (\bibinfo {year} {1970})}\BibitemShut {NoStop}%
\bibitem [{\citenamefont {Berger}(1972)}]{Berger_Phys_Rev_B_5_1862_1972}%
  \BibitemOpen
  \bibfield  {author} {\bibinfo {author} {\bibfnamefont {L.}~\bibnamefont
  {Berger}},\ }\bibfield  {title} {\bibinfo {title} {{Application of the
  Side-Jump Model to the Hall Effect and Nernst Effect in Ferromagnets}},\
  }\href {https://doi.org/10.1103/PhysRevB.5.1862} {\bibfield  {journal}
  {\bibinfo  {journal} {Phys. Rev. B}\ }\textbf {\bibinfo {volume} {5}},\
  \bibinfo {pages} {1862} (\bibinfo {year} {1972})}\BibitemShut {NoStop}%
\bibitem [{\citenamefont {Thouless}\ \emph {et~al.}(1982)\citenamefont
  {Thouless}, \citenamefont {Kohmoto}, \citenamefont {Nightingale},\ and\
  \citenamefont {den Nijs}}]{Thouless_Phys_Rev_Lett_49_405_1982}%
  \BibitemOpen
  \bibfield  {author} {\bibinfo {author} {\bibfnamefont {D.~J.}\ \bibnamefont
  {Thouless}}, \bibinfo {author} {\bibfnamefont {M.}~\bibnamefont {Kohmoto}},
  \bibinfo {author} {\bibfnamefont {M.~P.}\ \bibnamefont {Nightingale}},\ and\
  \bibinfo {author} {\bibfnamefont {M.}~\bibnamefont {den Nijs}},\ }\bibfield
  {title} {\bibinfo {title} {{Quantized Hall Conductance in a Two-Dimensional
  Periodic Potential}},\ }\href {https://doi.org/10.1103/PhysRevLett.49.405}
  {\bibfield  {journal} {\bibinfo  {journal} {Phys. Rev. Lett.}\ }\textbf
  {\bibinfo {volume} {49}},\ \bibinfo {pages} {405} (\bibinfo {year}
  {1982})}\BibitemShut {NoStop}%
\bibitem [{\citenamefont {Yao}\ \emph {et~al.}(2004)\citenamefont {Yao},
  \citenamefont {Kleinman}, \citenamefont {MacDonald}, \citenamefont {Sinova},
  \citenamefont {Jungwirth}, \citenamefont {Wang}, \citenamefont {Wang},\ and\
  \citenamefont {Niu}}]{Yao_Phys_Rev_Lett_92_037204_2004}%
  \BibitemOpen
  \bibfield  {author} {\bibinfo {author} {\bibfnamefont {Y.}~\bibnamefont
  {Yao}}, \bibinfo {author} {\bibfnamefont {L.}~\bibnamefont {Kleinman}},
  \bibinfo {author} {\bibfnamefont {A.~H.}\ \bibnamefont {MacDonald}}, \bibinfo
  {author} {\bibfnamefont {J.}~\bibnamefont {Sinova}}, \bibinfo {author}
  {\bibfnamefont {T.}~\bibnamefont {Jungwirth}}, \bibinfo {author}
  {\bibfnamefont {D.-s.}\ \bibnamefont {Wang}}, \bibinfo {author}
  {\bibfnamefont {E.}~\bibnamefont {Wang}},\ and\ \bibinfo {author}
  {\bibfnamefont {Q.}~\bibnamefont {Niu}},\ }\bibfield  {title} {\bibinfo
  {title} {{First Principles Calculation of Anomalous Hall Conductivity in
  Ferromagnetic bcc Fe}},\ }\href
  {https://doi.org/10.1103/PhysRevLett.92.037204} {\bibfield  {journal}
  {\bibinfo  {journal} {Phys. Rev. Lett.}\ }\textbf {\bibinfo {volume} {92}},\
  \bibinfo {pages} {037204} (\bibinfo {year} {2004})}\BibitemShut {NoStop}%
\bibitem [{\citenamefont {Klitzing}\ \emph {et~al.}(1980)\citenamefont
  {Klitzing}, \citenamefont {Dorda},\ and\ \citenamefont
  {Pepper}}]{Klitzing_Phys_Rev_Lett_45_494_1980}%
  \BibitemOpen
  \bibfield  {author} {\bibinfo {author} {\bibfnamefont {K.~v.}\ \bibnamefont
  {Klitzing}}, \bibinfo {author} {\bibfnamefont {G.}~\bibnamefont {Dorda}},\
  and\ \bibinfo {author} {\bibfnamefont {M.}~\bibnamefont {Pepper}},\
  }\bibfield  {title} {\bibinfo {title} {{New Method for High-Accuracy
  Determination of the Fine-Structure Constant Based on Quantized Hall
  Resistance}},\ }\href {https://doi.org/10.1103/PhysRevLett.45.494} {\bibfield
   {journal} {\bibinfo  {journal} {Phys. Rev. Lett.}\ }\textbf {\bibinfo
  {volume} {45}},\ \bibinfo {pages} {494} (\bibinfo {year} {1980})}\BibitemShut
  {NoStop}%
\end{thebibliography}%

\end{document}